\documentstyle[12pt]{article}

\newcommand{\definition}[2]{
\begin{figure}[htb]
\vspace{#1mm}
\begin{center}
\setlength{\unitlength}{0.6mm}
{\mbox{\begin{picture}(102,100)(0,0)
\thinlines
\put(-40, 30){\vector(0,1){10}}
\put(-40, 30){\vector(1,0){12}}
\put(-40, 30){\vector(2,3){4}}
\put(-38, 40){\makebox(0,0){$t$}}
\put(-26, 30){\makebox(0,0){$x$}}
\put(-32 ,36){\makebox(0,0){$y$}}
\put(-9, 30){\line(1,0){120}}
\put(-9, 90){\line(1,0){120}}
\put(115, 28){\makebox(0,0)[bl]{$t_0$}}
\put(115, 88){\makebox(0,0)[bl]{$t_1$}}
\thicklines
\put(1,30){\line(0,1){60}}
\put(16,30){\line(0,1){60}}
\put(46,30){\line(1,2){30}}
\put(61,30){\line(0,1){28}}
\put(61,62){\line(0,1){28}}
\put(101,30){\line(0,1){60}}
\put(31,60){\circle*{1}}
\put(41,60){\circle*{1}}
\put(71,60){\circle*{1}}
\put(81,60){\circle*{1}}
\put(91,60){\circle*{1}}
\put(0,20){\makebox(0,0)[bl]{1}}
\put(15,20){\makebox(0,0)[bl]{2}}
\put(45,20){\makebox(0,0)[bl]{$i$}}
\put(58,20){\makebox(0,0)[bl]{$i+1$}}
\put(100,20){\makebox(0,0)[bl]{$N$}}
\parbox{16cm}{\small #2}
\end{picture}}}
\end{center}
\end{figure}}

\newcommand{\ijji}[2]{
\begin{figure}[htb]
\vspace{#1mm}
\begin{center}
\setlength{\unitlength}{0.6mm}
{\mbox{\begin{picture}(202,100)(0,0)
\thinlines
\put(-20,30){\line(1,0){90}}
\put(-20,60){\line(1,0){90}}
\put(-20,90){\line(1,0){90}}
\put(110,30){\line(1,0){90}}
\put(110,60){\line(1,0){90}}
\put(110,90){\line(1,0){90}}
\put(74, 28){\makebox(0,0)[bl]{$t_0$}}
\put(74, 58){\makebox(0,0)[bl]{$t_1$}}
\put(74, 88){\makebox(0,0)[bl]{$t_2$}}
\put(204, 28){\makebox(0,0)[bl]{$t_0$}}
\put(204, 58){\makebox(0,0)[bl]{$t_1$}}
\put(204, 88){\makebox(0,0)[bl]{$t_2$}}
\thicklines
\put(-10,30){\line(0,1){60}}
\put(-5,30){\line(0,1){60}}
\put(5,30){\line(1,3){10}}
\put(10,30){\line(0,1){12}}
\put(10,48){\line(0,1){12}}
\put(25,30){\line(0,1){30}}
\put(30,30){\line(0,1){30}}
\put(60,30){\line(0,1){60}}
\put(10,60){\line(0,1){30}}
\put(15,60){\line(0,1){30}}
\put(25,60){\line(1,3){10}}
\put(30,60){\line(0,1){12}}
\put(30,78){\line(0,1){12}}
\put(120,30){\line(0,1){60}}
\put(125,30){\line(0,1){60}}
\put(135,30){\line(0,1){30}}
\put(140,30){\line(0,1){30}}
\put(165,30){\line(1,3){10}}
\put(170,30){\line(0,1){12}}
\put(170,48){\line(0,1){12}}
\put(190,30){\line(0,1){60}}
\put(135,60){\line(1,3){10}}
\put(140,60){\line(0,1){12}}
\put(140,78){\line(0,1){12}}
\put(170,60){\line(0,1){30}}
\put(175,60){\line(0,1){30}}
\put(-1,45){\circle*{1}}
\put(3,45){\circle*{1}}
\put(6,45){\circle*{1}}
\put(15,45){\circle*{1}}
\put(20,45){\circle*{1}}
\put(42,45){\circle*{1}}
\put(47,45){\circle*{1}}
\put(52,45){\circle*{1}}
\put(-1,75){\circle*{1}}
\put(3,75){\circle*{1}}
\put(6,75){\circle*{1}}
\put(20,75){\circle*{1}}
\put(25,75){\circle*{1}}
\put(42,75){\circle*{1}}
\put(47,75){\circle*{1}}
\put(52,75){\circle*{1}}
\put(128,45){\circle*{1}}
\put(132,45){\circle*{1}}
\put(147,45){\circle*{1}}
\put(153,45){\circle*{1}}
\put(159,45){\circle*{1}}
\put(176,45){\circle*{1}}
\put(180,45){\circle*{1}}
\put(184,45){\circle*{1}}
\put(130,75){\circle*{1}}
\put(135,75){\circle*{1}}
\put(147,75){\circle*{1}}
\put(153,75){\circle*{1}}
\put(159,75){\circle*{1}}
\put(180,75){\circle*{1}}
\put(185,75){\circle*{1}}
\put(91, 59){\line(1,0){8}}
\put(91, 61){\line(1,0){8}}
\put(4, 20){\makebox(0,0)[bl]{\small $i$}}
\put(9, 20){\makebox(0,0)[bl]{\small $i+1$}}
\put(24, 20){\makebox(0,0)[bl]{\small $j$}}
\put(29, 20){\makebox(0,0)[bl]{\small $j+1$}}
\put(134, 20){\makebox(0,0)[bl]{\small $i$}}
\put(139, 20){\makebox(0,0)[bl]{\small $i+1$}}
\put(164, 20){\makebox(0,0)[bl]{\small $j$}}
\put(169, 20){\makebox(0,0)[bl]{\small $j+1$}}
\parbox{16cm}{\small #2}
\end{picture}}}
\end{center}
\label{fig: ijji}
\end{figure}}

\newcommand{\fff}[2]{
\begin{figure}[hbt]
\vspace{#1mm}
\begin{center}
\setlength{\unitlength}{0.6mm}
{\mbox{\begin{picture}(240,150)(0,0)
\thinlines
\put(10,140){\line(1,0){80}}
\put(10,110){\line(1,0){80}}
\put(10,60){\line(1,0){80}}
\put(10,30){\line(1,0){80}}
\put(150,140){\line(1,0){80}}
\put(150,110){\line(1,0){80}}
\put(150,60){\line(1,0){80}}
\put(150,30){\line(1,0){80}}
\thicklines
\put(20,30){\line(3,4){60}}
\put(80,110){\line(0,1){30}}
\put(35,110){\line(1,1){30}}
\put(35,30){\line(0,1){18}}
\put(35,52){\line(0,1){58}}
\put(50,30){\line(0,1){38}}
\put(50,72){\line(0,1){51}}
\put(50,127){\line(0,1){13}}
\put(19,25){\makebox(0,0)[bl]{$a$}}
\put(34,25){\makebox(0,0)[bl]{$b$}}
\put(49,25){\makebox(0,0)[bl]{$c$}}
\put(49,143){\makebox(0,0)[bl]{$c$}}
\put(64,143){\makebox(0,0)[bl]{$b$}}
\put(79,143){\makebox(0,0)[bl]{$a$}}
\put(95,29){\makebox(0,0)[bl]{$t_0$}}
\put(95,59){\makebox(0,0)[bl]{$t_1$}}
\put(95,109){\makebox(0,0)[bl]{$t_2$}}
\put(95,139){\makebox(0,0)[bl]{$t_3$}}
\put(160,60){\line(3,4){60}}
\put(160,30){\line(0,1){30}}
\put(175,30){\line(1,1){30}}
\put(205,122){\line(0,1){18}}
\put(205,60){\line(0,1){58}}
\put(190,30){\line(0,1){13}}
\put(190,47){\line(0,1){51}}
\put(190,102){\line(0,1){38}}
\put(159,25){\makebox(0,0)[bl]{$a$}}
\put(174,25){\makebox(0,0)[bl]{$b$}}
\put(189,25){\makebox(0,0)[bl]{$c$}}
\put(189,143){\makebox(0,0)[bl]{$c$}}
\put(204,143){\makebox(0,0)[bl]{$b$}}
\put(219,143){\makebox(0,0)[bl]{$a$}}
\put(235,29){\makebox(0,0)[bl]{$t_0$}}
\put(235,59){\makebox(0,0)[bl]{$t_1$}}
\put(235,109){\makebox(0,0)[bl]{$t_2$}}
\put(235,139){\makebox(0,0)[bl]{$t_3$}}
\put(119,80){\line(1,0){13}}
\put(119,84){\line(1,0){13}}
\parbox{16cm}{\small #2}
\end{picture}}}
\end{center}
\end{figure}}

\newcommand{\four}[2]{
\begin{figure}[htb]
\vspace{#1mm}
\begin{center}
\setlength{\unitlength}{1mm}
{\mbox{\begin{picture}(120,40)(0,0)
\thicklines
\put(0, 30){\circle*{2}}
\put(10, 30){\circle*{2}}
\put(40, 30){\oval(10, 10)[bl]}
\put(40, 30){\oval(10, 10)[br]}
\put(40, 30){\circle*{2}}
\put(60, 30){\oval(10, 10)[bl]}
\put(60, 30){\oval(10, 10)[br]}
\put(60, 30){\circle*{2}}
\put(90, 30){\circle*{2}}
\put(40, 25){\vector(1, 0){1}}
\put(60, 25){\vector(1, 0){1}}
\thinlines
\put(-10, 30){\vector(1 ,0){110}}
\put(102, 30){\makebox(0,0){$x$}}
\put(0, 34){\makebox(0,0)[b]{1}}
\put(10, 34){\makebox(0,0)[b]{$2$}}
\put(40, 34){\makebox(0,0)[b]{$i+1$}}
\put(60, 34){\makebox(0,0)[b]{$j+1$}}
\put(90, 34){\makebox(0,0)[b]{$N$}}
\put(40, 22){\makebox(0,0){$t_0 \rightarrow t_1$}}
\put(60, 22){\makebox(0,0){$t_1 \rightarrow t_2$}}
\put(90, 22){\makebox(0,0){l.h.s of Figure 2}}
\put(40, 12){\makebox(0,0){$t_1 \rightarrow t_2$}}
\put(60, 12){\makebox(0,0){$t_0 \rightarrow t_1$}}
\put(90, 12){\makebox(0,0){r.h.s of Figure 2}}
\parbox{16cm}{\small #2}
\end{picture}}}
\end{center}
\end{figure}}

\newcommand{\gap}[2]{
\begin{figure}[htb]
\vspace{#1mm}
\begin{center}
\setlength{\unitlength}{0.6mm}
{\mbox{\begin{picture}(100,100)(0,0)
\thicklines
\put(0, 100){\line(1,0){100}}
\put(0, 20){\line(1,0){100}}
\thinlines
\put(20, 30){\line(1,0){60}}
\put(20, 40){\line(1,0){60}}
\put(50,60){\circle*{1}}
\put(50,65){\circle*{1}}
\put(50,70){\circle*{1}}
\put(50,75){\circle*{1}}
\put(50,80){\circle*{1}}
\put(105, 100){\makebox(0,0)[bl]{$n+1$}}
\put(105, 20){\makebox(0,0)[bl]{$n$}}
\put(85, 30){\makebox(0,0)[bl]{$m = 0$}}
\put(85, 40){\makebox(0,0)[bl]{$m = 1$}}
\parbox{8cm}{\small #2}
\end{picture}}}
\end{center}
\end{figure}}

\newlength{\extraspace}
\newlength{\extraspaces}
\newcommand{\beq}{\begin{equation}
\addtolength{\abovedisplayskip}{\extraspaces}
\addtolength{\belowdisplayskip}{\extraspaces}
\addtolength{\abovedisplayshortskip}{\extraspace}
\addtolength{\belowdisplayshortskip}{\extraspace}}
\newcommand{\eeq}{\end{equation}}

\newcommand{\beqa}{\begin{eqnarray}
\addtolength{\abovedisplayskip}{\extraspaces}
\addtolength{\belowdisplayskip}{\extraspaces}
\addtolength{\abovedisplayshortskip}{\extraspace}
\addtolength{\belowdisplayshortskip}{\extraspace}}
\newcommand{\eeqa}{\end{eqnarray}}

\newcommand{\newsection}[1]{
\vspace{10mm}
\pagebreak[3]
\addtocounter{section}{1}
\setcounter{equation}{0}
\setcounter{subsection}{0}
\setcounter{footnote}{0}
\begin{center}
{\large\bf \thesection. #1}
\end{center}
\nopagebreak
\medskip
\nopagebreak}

\newcommand{\newsubsection}[1]{
\vspace{5mm}
\pagebreak[3]
\addtocounter{subsection}{1}
\noindent{ \sc \thesubsection. #1}
\nopagebreak
\smallskip
\nopagebreak}

\newcommand {\dell}[1]{\frac{\partial}{\partial #1}}
\newcommand {\delldell}[2]{\frac{\partial #1}{\partial #2}}
\newcommand{\fpar}{\hspace{6mm}}
\newcommand{\fr}[1]{{1 \over{#1}}}
\newcommand{\ie}{{\it i.e}.\ }

\newcommand{\is}{\! & \! = \! & \!}
\newcommand{\nonu}{\nonumber \\[1.5mm]}
\newcommand{\half}{{1 \over{2}}}

\begin{document}
\addtolength{\baselineskip}{.7mm}

\begin{flushright}
{\sc NUS/HEP/}92011\\
January 1991
\end{flushright}
\vspace{.3cm}

\begin{center}
{\Large{\bf{Spinning Braid Group Representation\\[2mm]
and the Fractional Quantum Hall Effect}}}\\[13mm]

{\sc Christopher Ting}\footnote{e-mail: scip1005@nusvm.bitnet.}\\[3mm]
{\it Defence Science Organization,
20 Science Park Drive, Singapore 0511} \\[.4cm]
{and}\\[.4cm]
{\it Department of Physics,
National University of Singapore\\[2mm]
Lower Kent Ridge Road, Singapore 0511} \\[.4cm]
{and}\\[.4cm]
{\sc C. H. Lai}\footnote{e-mail: phylaich@nusvm.bitnet.}\\[3mm]
{\it Department of Physics,
National University of Singapore\\[2mm]
Lower Kent Ridge Road, Singapore 0511}
\\[8mm]
{\bf Abstract}
\end{center}
\noindent
The path integral approach to representing braid group is generalized for
particles with spin. Introducing the notion of {\em charged}
winding number in the super-plane, we represent the braid group generators
as homotopically constrained Feynman kernels. In this framework,
super Knizhnik-Zamolodchikov operators appear naturally in
the Hamiltonian, suggesting the possibility of {\em spinning
nonabelian} anyons. We then apply
our formulation to the study of fractional quantum Hall effect (FQHE).
A systematic discussion of the ground states and their quasi-hole
excitations
is given. We obtain Laughlin, Halperin and Moore-Read states as
{\em exact} ground state solutions to the respective Hamiltonians
associated to the braid group representations.
The energy gap of the quasi-excitation is also obtainable from this approach.
\vfil
\newpage
\newsection{Introduction}
\fpar
The fractional quantum Hall effect (FQHE) \cite{Tsui et al} is a
collective phenomenon of $N$ electrons living in an effectively 2-dimensional
plane. Under suitable conditions, the Hall conductivity is ``quantized" as
${p\over{q}} \, {e^{2}\over{h c}}$, \ie plateaux pegged at these
values for some integers $p$ and $q$ are observable
along the axis of the strength $B$ of the external magnetic field. This
macroscopic quantum behaviour has been successfully captured by Laughlin's
theory when $p=1$ and $q$ is an odd number \cite{Laughlin}.
Essentially, the ground states are that of an
incompressible quantum liquid. The particle-like excitations,
called quasi-particles and quasi-holes respectively, are some
gap away from the ground state in the spectrum. They do not contribute to
the transport coefficients because of localization effect. What is more
interesting is that they are fractionally charged anyons. The reason why the
filling fractions have odd denominators is that the many-body ground states
proposed by Laughlin must pick up a minus sign whenever any two electrons
swap positions; afterall, electrons are fermions. These Laughlin states
constitute the
corner stones of the theory of FQHE. All the key ingredients such as
the existence of a finite energy gap, and the fractional statistics
of quasi-excitations follow from the ans\"{a}tz.

Nevertheless, nature vouchsafes more pleasant surprises.
FQHE with even-denominator filling fractions was discovered
\cite{Willett et al}\cite{Eisenstein et al}. If not for
this discovery, Laughlin's theory would have been more or
less adequate\footnote{Additional ideas, though, are needed to
account for those plateaux with $p \neq 1$. At any rate, it is not
unfair to say that they all build
upon the conceptual foundation of Laughlin's theory.}.
To account for these even filling fractions,
spin-unpolarized states have been proposed
\cite{HR}. Despite some
disagreements, the consensus is that the electron's spin,
which is totally frozen out in Laughlin's picture
plays a role in the occurrence of even-denominator states.
Experimental evidence of an unpolarized state even for odd
denominator filling fraction \cite{8/5} makes it all the more
imperative to scrutinize the role of spin in FQHE.

With this in mind, we propose here a microscopic $N$-body
Hamiltonian obtained from the path integral representation of
the braid group \cite{Lai-Ting}. When this Hamiltonian is minimally
coupled to the background gauge potential of the uniform external
magnetic field in the symmetric gauge,
one finds that Laughlin states are {\em exact} ground states.
This was done for electrons carrying the representation of
$U(1)$ \cite{Ting-Lai}.
Furthermore, since our formulation is a non-abelian generalization of
Y. S. Wu's \cite{Wu}, it is possible to
proceed directly to consider the case where
the representation carried by the electrons is $SU(2) \times U(1)$;
presumably, spin may be regarded as isospin in the non-relativistic
regime \cite{spin}.
Thus, we obtain Halperin state as {\em exact} ground state solution.
We also extend our previous works by switching on the spin
degree of freedom in an alternative fashion.
Using Grassmannian variables to formulate
spin as dynamical variable, we study
the path integral of free spinning particles on a super-plane.
In this manner,
we obtain non-trivial results generalizing the spinless case.
With this approach, we get a wavefunction which is the exact ground state
of the spinning Hamiltonian in the external
magnetic field. It turns out to be
the same as the one constructed by Moore and Read
which is a product of
some conformal blocks of the Ising model and rational torus
\cite{MR}.
{}From these analyses, we conclude that
(super-)Knizhnik-Zamolodchikov
operator minimally coupled to the background gauge field
is the {\em microscopic} ground state equation of FQHE.

In section 2, we review the basic ideas leading to the
path integral representation of braid group. After proposing
the quantization procedure for the {\em spinning} quantum mechanics,
we proceed to construct an analogous representation with the path integral
of free particles with spin in section 3. We
then consider the link between braid group statistics and FQHE. The
ground state equations are solved for polarized FQHE states,
followed by the spinning states in section 4. Since the
key issue of FQHE is its incompressibility, we feature the topological
origin of the quasi-excitations in section 5 and
suggest how the energy gap may be obtained
from this perspective. In particular, a novel formula to calculate the
ratio of the energy gaps of Laughlin states is presented.
In section 6, we see how Halperin's state is obtained as the exact solution
of the ground state equation of spin singlets. In section 7,
we discuss the connection with WZW models and the crucial role played
by the external magnetic field. The main results are summarized in
the last section.

\newsection{Path Integral Representation}
\fpar
Artin's braid group $B_N$ is intrinsically a 3-dimensional object
which comprises of $N$ ambient isotopic classes of curves in
${\bf R}^3$. An intuitive
representation of the elements of the group
is to use a number of threads and weave them.
Given $N$ threads, the elements
of $B_N$ can be constructed from $N-1$ basic weaves
$\sigma_i$, $i = 1, \cdots\cdots N-1$. Here $\sigma_i$
is used to denote
a pattern in Figure 1, where the $i$-th thread crosses
{\em over} the $i+1$-th thread.
It is worth remarking that the pictorial representation of $\sigma_i$
is faithful and irreducible.
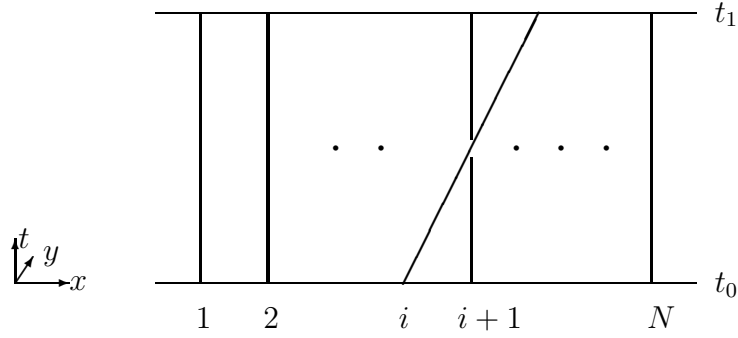
\definition{0}{\hspace{10mm} Fig. 1: The action of $\sigma_i$.}

These braid group generators
satisfy the following algebraic relations:
\beq
        \sigma_{i} \sigma_{j} = \sigma_{j} \sigma_{i}, \hspace{20mm}
           |i - j| \geq 2,
        \label{eqn: ijji}
\eeq
\beq
        \sigma_{i} \sigma_{i+1} \sigma_{i}
                  =  \sigma_{i+1} \sigma_{i} \sigma_{i+1}, \hspace{10mm}
           i = 1, \cdots, n - 2.
        \label{eqn: ad}
\eeq
The word $\sigma_i \sigma_j$, for instance,
has been represented as putting
one diagram on top of the other
as shown in the
left-hand side of Figure 2.
The meanings of (\ref{eqn: ijji}) and
(\ref{eqn: ad}) are explicit from the weave
patterns depicted in Figure 2 and Figure 3 respectively.
When stacking the diagrams, one has to exchange the
labels to ensure that the glued world lines
carry the same representations throughout. This is implicitly
carried out in the figures.
\ijji{0}{\hspace{4.5cm} Fig. 2: $\sigma_i \sigma_j = \sigma_j \sigma_i$.}
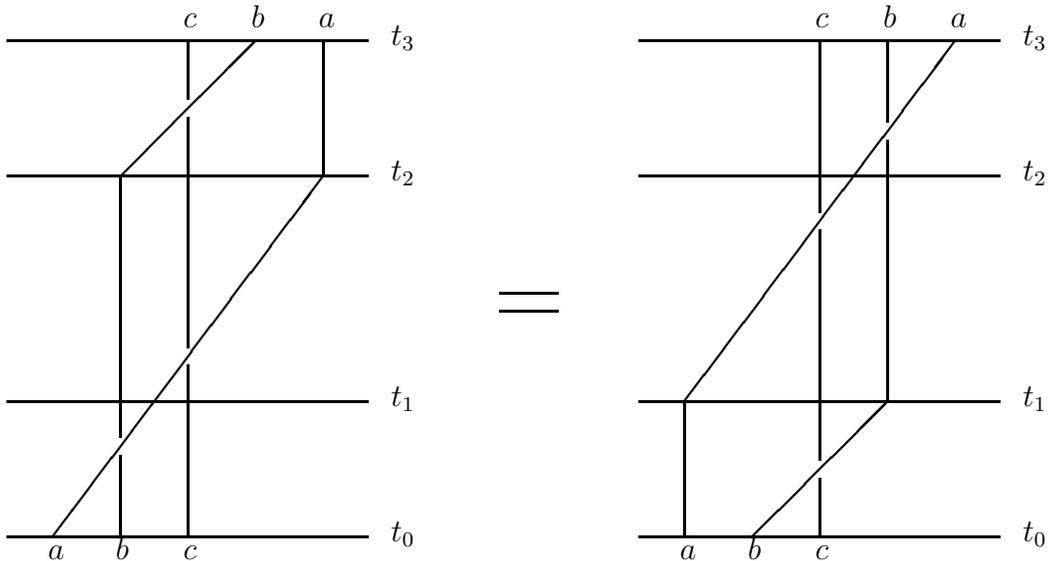
\fff{0}{\hspace{2cm} Fig. 3: Graphical representation of
$\sigma_i\sigma_{i+1}\sigma_i=\sigma_{i+1}\sigma_i\sigma_{i+1}$.}
\par
The basic idea of the path integral representation \cite{Wu} is to see
the threads as non-relativistic world lines of point particles. In this
light, the path parametrized by time $t$ of $i$-th particle in the
2-dimensional plane is
conceivable. By definition, the number of threads $N$ is a constant
of motion; at all times, no two threads can fuse together and become one.
In the language of the configuration space $M_N$ of $N$ particles, it
means that the topology is multiply connected. Each particle sees the rest
as punctures. As opposed to higher dimensions, the fundamental
group of the configuration space $\pi_1(M_N)$ is an infinite
non-abelian group, by virtue of which particles that are neither
bosons nor fermions are theoretically allowed.
It turns out that $\pi_1(M_N)$ is isomorphic to the pure braid group if
all the particles do not
carry the same representation, and $\pi_1(M_N) \cong B_N$
if they do.

Because the configuration
space is multiply connected, the paths are homotopically
classified and those of different classes cannot be smoothly
deformed from one to the other. When one considers the
Feynman kernel for a particle to move from point $z_{a(0)}$
at time $t_0$ to point $z_{a(1)}$ at time $t_1$,
one has to organize the paths according to their homotopical
classes. Now, the homotopy class of a path in $M_N$
is determined by the winding numbers with respect to the
punctures. The number of times a
path goes around a puncture is well-defined and non-trivial
only when the path is in 2-dimensional space.
It is this peculiarity of the spatial dimension being two that gives
rise to the possibility of anyonic statistics.

Earlier, we have generalized these ideas to particles carrying
representations
of a non-abelian group \cite{Lai-Ting}. We shall briefly review the
work here.
To construct a non-abelian representation of the braid group,
we introduce the notion of {\em charged} winding number $w$ for a path:
\beq
      w = \frac{1}{2\pi i} \int_{C} \frac{dz_a}{z_a - z_b}
                                  T_a \otimes T_b.
\label{windnum}
\eeq
Here,
$T_a$ and $T_b$ are the representations carried by the
particles. In this manner, the threads are
more than merely worldlines; they have become Wilson lines. The charged
winding angle $\Theta$ can now be defined as:
\beq
       \Theta = \hbox{sign}(C)~|\Theta_{a(1)} - \Theta_{a(0)}| + 2\pi w.
       \label{wind}
\eeq
We choose the convention that a path going counterclockwise about
the puncture
$z_b$ has positive sign, namely $\hbox{sign}(C) = 1$, and
denote $\vartheta = \hbox{sign}(C)~|\Theta_{a(1)} - \Theta_{a(0)}|$.
With this convention,
for the homotopically equivalent paths corresponding to
$\sigma_i$, which cross over from the left, the
change in the azimuthal angle $\vartheta$ is
non-negative.

The constrained Feynman kernel of homotopy class $l$ for particle
$a$ with mass $m$ can be expressed formally as:
\beq
     K_{l}(z_{a(1)}, t_{1}, z_{a(0)}, t_{0}) =
       \int {\cal D}_{l}z_a(t) {\cal D}_{l}\overline{z}_a(t)
       \exp~i\int_{t_{0}}^{t_{1}}\frac{1}{2} m |\dot{z}_a(t)|^2 dt~~
       \delta^2 ( 2\pi l~T_{a} \otimes T_{b}- \Theta ).
       \label{Fk}
\eeq
With the path ordering determined by that in the definition
of charged winding angle
(\ref{wind}), the matrix-valued Dirac delta function can be
represented by the following path-ordered Fourier transform:
\beq
   \delta^2 ( 2\pi l~T_{a} \otimes T_{b}- \Theta ) = \int\!\!\int
                \frac{dk}{2\pi} \frac{d\overline{k}}{2\pi}\,
                e^{-i(k\vartheta + {\overline k}\vartheta )}
          \,\hbox{P} \exp~i\left[(2\pi k(l~T_{a} \otimes T_{b} - w)
                + c.c.\right].
          \label{delta_function}
\eeq
This expression is nothing but a functional integral description
of the topological properties of the configuration space.
It is the main ingredient of our representation.
Technically, the way we formulate
the homotopic constraint via (\ref{delta_function})
is quite different from Wu's \cite{Wu}.
Here, the change of azimuthal angle $\vartheta$
is fixed by the initial and final
positions of particle $a$. Substituting (\ref{delta_function})
into the Feynman kernel (\ref{Fk}), we obtain the Fourier
transform:
\beq
     K_{l}(z_{a(1)}, t_{1}, z_{a(0)}, t_{0}) =
        \int\!\!\int \frac{dk}{2\pi} \frac{d\overline{k}}{2\pi}~
                e^{-i(k\vartheta + \overline{k}\vartheta )}\,
                \widetilde{K}_{l}(z_{a(1)}, t_{1}, z_{a(0)},
                 t_{0}; k, {\overline k}),
    \label{Ft}
\eeq
where
\beqa
   \widetilde{K}_{l}(z_{a(1)},t_{1},z_{a(0)},t_{0};k,\overline{k})
   \! &=& \!
   \int {\cal D}_{l}z_a(t) {\cal D}_{l}\overline{z}_a(t)\,
   \hbox{P}\! \exp \, i\int_{t_{0}}^{t_{1}}
   \frac{1}{2} m |\dot{z}_a(t)|^2~dt \nonu
   &\quad & \times\, \exp \, i \, \int_{t_{0}}^{t_{1}} \left(
   k( \frac{ i \, \dot{z}_a}{z_a - z_b} + 2\pi  l )
   T_{a} \otimes T_{b} + c.c. \right)~dt.\nonu
   \label{tildeFk}
\eeqa

Expressions (\ref{Fk}) and (\ref{tildeFk}) can be easily generalized
to $N$ particles at $z_1, z_2, \cdots , z_N$, with
$\mbox{\rm Re} \, z_1 < \mbox{\rm Re} \, z_2
< \cdots < \mbox{\rm Re} \, z_N$.
Let particle $i$ make a trip from $z_{i \, (0)} = z_{i}(t_{0}) = z_{i}$ to
$z_{i \, (1)} = z_{i}(t_{1}), \, \mbox{\rm Re}
\, z_{i \, (1)} > \mbox{\rm Re} \, z_{i+1}.$
Denoting the difference in the initial angle and the
final angle of the paths of particle $i$ with respect to particle $j$ as
$\vartheta_{ij}$,
$\vartheta_{ij} =\hbox{sign}(C_{i})~|\Theta_{ij \, (1)} -
\Theta_{ij \, (0)}|$,
the constrained Feynman kernel of
homotopy class $( l_{1},\cdot\cdot, l_{i-1}, l_{i+1}, \cdot\cdot, l_{n})$ for
particle $i$ carrying representation $T_i$ is
\beqa
    & & K_{l_{i}}(z_{i \, (1)}, t_{1}, z_{i \, (0)}, t_{0}) = \nonu
    & &\qquad\int\!\!\int \frac{dk}{2\pi} \frac{d\overline{k}}{2\pi}~
                \exp \left(-i \sum_{j =1, j \ne i}^n
                       \left( k\vartheta_{ij} + \overline{k}\vartheta_{ij}
                       \right) \right)
       \widetilde{K}_{l_{i}}(z_{i \, (1)}, t_{1}, z_{i \, (0)}, t_{0}; k,
       \overline{k}),
\eeqa
where
\beqa
     \widetilde{K}_{l_{i}}(z_{i\, (1)},t_{1},
     z_{i\, (0)},t_{0};k,\overline{k}) &=&
     \int {\cal D}_{l_{i}}z_{i}(t) {\cal D}_{l_{i}}\overline{z}_{i}(t)\,
     \hbox{P}\,\exp~i\int_{t_{0}}^{t_{1}}
     \frac{1}{2} m_{i} |\dot{z}_{i}(t)|^2~dt  \nonu
     &\quad &\times \exp \, i \, \int_{t_{0}}^{t_{1}} \left(
     k \sum_{j = 1, j \ne i}^n
     \left( \frac{ i \, \dot{z}_{i}}{z_{i} - z_{j}} + 2\pi l_{j} \right)
     T_{i} \otimes T_{j} + c.c. \right) \, dt. \nonu
\eeqa
Given these initial and final conditions, $\sigma_{i}$ can be represented by
the positively oriented Feynman kernel of class
$(0, \cdot\cdot,\widehat{0}_{i}, \cdot\cdot, 0),$ the $i$-th 0
is omitted as we do not consider self-linking. The self-linking
problem does not arise here because Feynman kernels are defined
for $t \geq 0$ only. Writing,
\beqa
                A_{z_{i}} \is T^{\alpha}_{i} A^{\alpha}_{z_{i}} =
                   i k \sum_{j=1, j \ne i}^N
                   \frac{ T_{i} \otimes T_{j}}{ z_{i} - z_{j}}, \\
                A_{\overline{z}_{i}} \is
                   \overline{T}^{\alpha}_{i} A^{\alpha}_{\overline{z}_{i}} =
                   i \overline{k} \sum_{j=1, j \ne i}^N
                   \frac{ \overline{T}_{i} \otimes \overline{T}_{j}}
                        { \overline{z}_{i} - \overline{z}_{j} },
\eeqa
the proposed representation $D(\sigma_{i})$ is
$K_{i}(t_{1},t_{0};\vartheta_{i\, 1},\cdot\cdot,\vartheta_{i\, i-1},
\vartheta_{i\, i+1},
\cdot\cdot, \vartheta_{i \, N})$ given below:
\beqa
       \int {\cal D}_{+}z_{i} {\cal D}_{+}\overline{z}_{i}
       \int\!\!\int \frac{dk}{2\pi} \frac{d\overline{k}}{2\pi}~\hbox{P}\,
       \exp~i \, \int_{C_{i}}
       \frac{1}{2} m_{i} |dz_{i}|^2
       + A_{z_{i}} dz_{i} + A_{\overline{z}_{i}} d\overline{z}_{i} \nonu
       \times\,\exp \left(-i \sum_{j =1, j \ne i}^N
       \left( k\vartheta_{ij} + \overline{k}\vartheta_{ij} \right)
       \right),
       \label{rep}
\eeqa
followed by an exchange operation $\Pi_{i \, i+1}$,
\beq
      D(\sigma_{i}) = \Pi_{i \, i+1} \,
  K_i(t_1,t_0;\vartheta_{i\, 1},\cdot\cdot,
  \vartheta_{i\, i-1},\vartheta_{i\,i+1}, \cdot\cdot, \vartheta_{i\, N}).
\eeq
$\Pi_{i \, i+1}$ is to make every
world line stick to the same representation space it has
started with.
The multiplication rule for braid group generators is realised as the usual
multiplication of kernels.

It remains to verify that
\beqa
        D(\sigma_i) D(\sigma_j) &=&  D(\sigma_j) D(\sigma_i), \hspace{27mm}
           |i - j| \geq 2, \\
        \label{eqn: Dadijji}
        D(\sigma_i) D(\sigma_{i+1}) D(\sigma_i)
                  &=&  D(\sigma_{i+1}) D(\sigma_i) D(\sigma_{i+1}),
        \hspace{10mm}  i = 1, \cdots, N - 2.
        \label{eqn: Dadfff}
\eeqa
One can first look at the paths in
the plane corresponding to the
space-time diagrams of Figure 2 and 3.
It is obvious that (\ref{eqn: Dadijji}) holds; the two paths
in the plane are disjoint by definition (Figure 4).
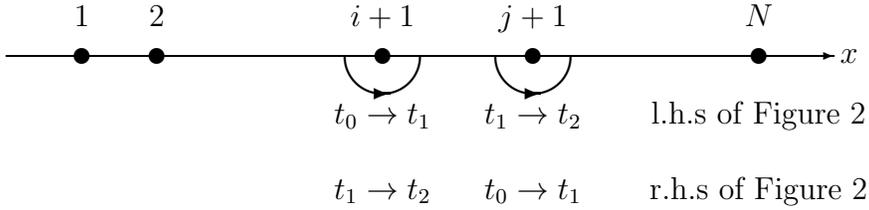
\four{0}{Fig 4: The paths of particles $i$ and $j$ in the $x$-$y$ plane.}
Similarly, the proof of (\ref{eqn: Dadfff}) is readily seen from
Figure 5.
Upon careful examination of the overall changes in the
azimuthal angles before ($t_0$) and after ($t_3$), one finds that
the two figures give the same results; it does not matter whether particle
$a$ moves first as in (A) of Figure 5 or particle
$b$ in (B). The expressions in
terms of Feynman kernels for the proof of (\ref{eqn: Dadfff}) were given
in \cite{Lai-Ting}.

Now, the effective Lagrangian of particle $i$ can be readily read from
(\ref{rep}).
\beq
    L = \frac{1}{2} m_{i} |\dot{z_{i}}|^2
        + A_{z_{i}} \dot{z}_{i} + A_{\overline{z}_{i}} \dot{\overline{z}}_{i}.
    \label{Lagrangian}
\eeq
It is amusing that $A_{z_i}, A_{\overline{z}_{i}}$, together with
$A_{0_i} = 0$ may be seen as the components of
some gauge field in the temporal guage. In fact,
$A^{\alpha}_{z_{i}}$ and $A^{\alpha}_{\overline{z}_{i}}$ satisfy
Gauss' law:
\beq
      \frac{k}{2 \pi} F^{\alpha}_{z_{i} \overline{z}_i} =
        - \sum_{j=1, j \neq i}^N T^{\alpha}_{j} \delta^{2}(z_{i} - z_{j}),
          \hspace{10mm} \alpha = 1, \cdots, {\rm dim~G},
      \label{Gauss}
\eeq
where $F^{\alpha}_{z_{i} \overline{z_{i}}}$
are the components of the field strength. In a sense, this result furnishes
an interpretation to Witten's Chern-Simons theory
\cite{Witten}\cite{CS-theory}\cite{GMM}:
The topological quantum field theory
of pure Chern-Simons action can be embedded in a non-relativistic,
quantum mechanical system of free particles. To see this, we consider
the Schr\"{o}dinger equation associated to the Feynman kernel of particle
$i$:
\beq
   i \frac{\partial}{\partial t} \psi = -\frac{1}{m_{i}}
   \left[ ( \partial_{z_{i}} - i A_{z_{i}} )
          ( \partial_{\overline{z}_{i}} - i A_{\overline{z}_{i}}) +
          ( \partial_{\overline{z}_{i}} - i A_{\overline{z}_{i}})
          ( \partial_{z_{i}} - i A_{z_{i}} ) \right] \psi.
  \label{Sch}
\eeq
In the limit $m_{i}\to 0,$ a
class of solutions of (\ref{Sch}) consists of those wavefunctions
$\psi$ satisfying
\beq
         ( \partial_{z_{i}} - i A_{z_{i}} )  \psi
         =  \left( \frac{\partial}{\partial z_{i}}
              + k \sum_{j=1, j \ne i}^N
                   \frac{ T_{i} \otimes T_{j}}{ z_{i} - z_{j}} \right) \psi
          =0 \, ,
         \label{KZ1}
\eeq
\beq
          ( \partial_{\overline{z}_{i}} - i A_{\overline{z}_{i}}) \psi
         = \left( \frac{\partial}{\partial \overline{z}_{i}}
             + \overline{k} \sum_{j=1, j \ne i}^N
                   \frac{ \overline{T}_{i} \otimes \overline{T}_{j}}
                        { \overline{z}_{i} - \overline{z}_{j} } \right) \psi
         = 0 \, .
        \label{KZ2}
\eeq
These are precisely the Knizhnik-Zamolodchikov equations if we set
$k = \overline{k}=-2/(l+c_V)$, where $l$ is the level of the WZW model
and $c_{V}$ is the quadratic Casimir of the adjoint representation
of the group $G$ \cite{KZ}. It is interesting to note that the
wavefunctions, though non-normalizable, are the parallel transport sections
of a complex vector bundle over the base manifold $M_N$.

In the context of particle
statistics, (\ref{Sch}) can be interpreted as the Schr\"odinger equation for
``non-Abelian'' anyons. When $T_i = \overline{T}_i = 1$, $i = 1, \cdots, N$,
it is the (abelian)
1-dimensional irreducible representation constructed by Wu
\cite{Wu}. Therefore our construction is a non-Abelian generalization
of the general theory of quantum statistics in two dimensions.

\newsection{Spinning Path Integral Representation}
\fpar
The braid group representation constructed in the previous
section can be generalized for particles with spin. In this section,
we first propose a {\em spinning} quantization rule suitable for
such purpose. The little difference with the usual supersymmetric
quantum mechanics is that the eigen-wavefunction of
the spinning
Hamiltonian describing the dynamics in the super plane can be found before
integrating out the anti-commuting axes $\theta, \overline{\theta}$.
Then, using the definition of a super winding number and its {\em charged}
version, we construct the spinning path integral representation of
Artin's braid group.

\newsubsection{Spinning Quantum Mechanics\footnote{In this subsection,
we set all the universal constants $\hbar = c = e = 1$, as well as the mass
of the particle and the magnetic field strength equal to 1.
}}
\par
The spin degree of freedom may be described in terms of the Grassmannian
variables. For a single spinning electron in the world of flat-land,
the configuration space is $\mbox{\bf R}^2 \times Gr_2$. The non-relativistic
quantum mechanics of a free, spinning particle in the flat-land can be
formulated as the sum over all possible paths in the super-plane. The real
commuting variables $x$ and $y$ denote the coordinates of the plane, and
$\theta, \overline{\theta}$ the anti-commuting ``axes" for the spin degree of
freedom. Now, the dynamical variables of the particle in the configuration
space can be specified by:
\beqa
\phi (t) \is z(t) + i \overline{\theta} \xi (t),  \nonu
\overline\phi (t) \is \overline{z}(t) + i \theta \overline\xi (t) \, ,
\eeqa
where $z = x + iy $, ${\overline z} = x - iy$ and
$\xi, \overline\xi$ are the Grassmannian variables for the components
of the spin degree of freedom in flat-land.
Thus, we regard the dynamical degrees
of freedom of the particle as a pair of
{\em chiral} superfields. (Quantum mechanics
can be seen as 1-dimensional ``field" theory,
the dimension being time $t$.)
Notice that $\phi$ and $\overline\phi$ have even Grassmannian parity. The
Lagrangian of a free spinning particle is
\beq
L = \half |\dot{z}|^2 + {i \over{2}}
( \overline\xi \dot\xi - \dot{\overline{\xi}} \xi ).
\label{L}
\eeq
In terms of superfields, we have
\beq
L = \int d\overline\theta \, d\theta \, {\cal L},
\eeq
where
\beq
{\cal L} = \half (\theta \dot{\overline{\phi}}) (\overline\theta \dot\phi)
+ {i\over{2}} (\dot\phi \overline\phi - \dot{\overline{\phi}} \phi).
\label{superL}
\eeq
In the calculation, we have adopted the following convention for
the Berezin integral:
\beqa
\int d \theta \is \int d \overline{\theta} \, = 0 \, , \nonu
\int d \theta \, \theta \is \int d \overline{\theta}
\, \overline{\theta} = 1 \, .
\eeqa
Though the form of the spinning Lagrangian is
exactly the same as (\ref{L}), there is a difference between them.
The first term in (\ref{superL}) is now a product of odd variables and the
second term is composed of even variables. This is the reverse of
(\ref{L}), where $z$ is even and $\xi, \overline{\xi}$ odd. The
spinning Hamiltonian can be obtained from the Legendre transformation,
with the canonical momenta
${\cal{P, \overline{P}}}, \pi, \overline{\pi}$
defined and calculated as follows.
\beqa
{\cal P}
\! & \! \equiv \! & \!
\delldell{{\cal L}}{(\theta \dot{\overline{\phi}})}
= \half \overline{\theta} \dot{\phi} \, , \nonu
\overline{\cal P}
\! & \! \equiv \! & \!
\delldell{{\cal L}}{({\overline{\theta}} \dot{\phi})}
= - \half \theta \dot{\overline{\phi}} \, , \nonu
\pi \! & \! \equiv \! & \!
\delldell{{\cal L}}{\dot{\phi}} = {i\over{2}} \overline{\phi} \, , \nonu
\overline{\pi} \! & \! \equiv \! & \!
\delldell{{\cal L}}{\dot{\overline{\phi}}} = - {i\over{2}} \phi \, .
\label{cm}
\eeqa
The minus sign of the second expression in (\ref{cm}) is a property of
the chain rule for differentiating a product of Grassmannian odd variables.
The consistency of the formulation can be checked by examining whether the
spinning Hamiltonian thus obtained reproduces the usual Hamiltonian after
integrating over $\theta$ and $\overline{\theta}$.
\beqa
{\cal H} \is ( \theta \dot{\overline{\phi}} ) \,
{\cal P}
+ ( \overline{\theta} \dot{\phi} ) \,
\overline{\cal P}
+ \dot{\phi} \, \pi
+ \dot{\overline{\phi}} \, \overline{\pi}
- {\cal L} \nonu
\is \half (\theta \dot{\overline{\phi}}) (\overline\theta \dot\phi).
\eeqa
Since
\beq
\int d \overline{\theta} d \theta \, {\cal H} = \half |\dot{z}|^2,
\eeq
we see that the spinning formulation is correct; in the absence of
magnetic field, the spin degree of freedom is hidden and the
energy spectrum of a free spinning particle
is determined exclusively by the kinetic energy.
Now we introduce the differential operators
whose Grassmannian parity is odd:
\beqa
D_z \equiv  \dell{\theta} + \theta \dell{z} \, , \nonu
D_{\overline{z}} \equiv  \dell{\overline{\theta}}
+ \overline{\theta} \dell{\overline{z}} \, .
\label{qr}
\eeqa
The usual quantization rule $[q_i \, , \, p_j] = i \delta_{ij}$ for pairs
of canonical variables $q_i, p_i, \, i = 1, 2, \cdots$
takes the following form in the spinning
formalism:
\beqa
\{ \overline{\theta} \phi \, , \, {\cal P} \} \is i \, \overline{\theta} \theta
= -i \, \theta\overline{\theta}\, , \nonu
\{ \theta \overline{\phi} \, , \, \overline{{\cal P}}  \}
\is i \, \theta \overline{\theta},
\eeqa
where $\{ \, , \, \}$ is anticommutator since all the operators
entering the bracket in (\ref{qr}) are odd.
Using the definitions of $\phi, \overline{\phi}$ and $D_z, D_{\overline z}$,
it is straightforward to calculate that
\beqa
\{ D_z, \,  \overline{\theta} \phi \} \is \theta \overline{\theta} \, , \nonu
\{ D_{\overline z}, \, \theta \overline{\phi} \}
\is - \theta \overline{\theta} \, ,
\eeqa
Therefore we can represent the coordinates and momenta operators as
$\overline{\theta} \phi \rightarrow \overline{\theta} \phi$,
$\theta \overline{\phi} \rightarrow \theta \overline{\phi}$,
${\cal P} \rightarrow -i D_z$, and
$\overline{\cal P} \rightarrow -i D_{\overline z}$. This representation is
the spinning analogue of the usual
Schr\"{o}dinger representation.

The spinning eigen-wavefunction
$\Psi$ of ${\cal H}$
can be defined with respect to the eigen-wavefunction $\psi$ of $H$
as follows.
\beqa
H \psi \is \left( \int d \overline{\theta} d \theta \, {\cal H} \right)
\psi \nonu
\! & \! \stackrel{\rm def}{=} \! & \!
         \int d \overline{\theta} d \theta \, {\cal H} \Psi.
\label{eq: spin eigen}
\eeqa
Since if $E$ is the eigenvalue of both H and $\cal H$, \ie
$H \psi = E \psi$ and ${\cal H} \Psi = E \Psi$, we have
\beq
\psi = \int d \overline{\theta} d \theta \, \Psi.
\label{psi}
\eeq

As an example of this formalism, let us consider the
quantum mechanics of a spin-$\half$
particle moving
in an external magnetic field which is
uniform, constant
and perpendicular to the plane.
The minimally coupled covariant derivatives
are
\beqa
{\cal D} \equiv  D_z + i \theta A_z \, , \nonu
\overline{{\cal D}} \equiv
D_{\overline z} + i \overline{\theta} A_{\overline z},
\eeqa
where
$B_z, B_{\overline z}$ denote the components of the
gauge field of the magnetic field.
In the symmetric guage $B_z = -i {{\overline z} \over 4}$,
$B_{\overline z} = i {z \over 4}$,
we have
\beqa
{\cal D} \is \dell{\theta}
+ \theta( \dell{z} + \frac{\overline{z}}{4} )\, , \nonu
\overline{\cal D} \is
\dell{\overline{\theta}} + \overline{\theta}
( \dell{\overline{z}} - \frac{z}{4} ).
\eeqa
One finds that
\beq
\{ {\cal D}, \, \overline{{\cal D}} \} = -  \frac{\theta\overline{\theta}}{2}.
\eeq
So in the Schr\"odinger representation
the spinning Hamiltonian is
\beqa
{\cal H} \is \overline{\cal D}{\cal D} - {\cal D}\overline{\cal D}
+ \half [\overline{\phi}, \phi] \nonu
\is 2 \overline{\cal D}{\cal D} + \half \theta\overline{\theta}
+ \half [\overline{\phi}, \phi],
\eeqa
$[ \, \, , \, ]$ being commutator.
The quantization rule for $\phi, \overline{\phi}$
before integrating out the Grassmannian axes is the usual one:
$\{ \overline{\phi} \, , \, \phi \} = \theta\overline{\theta}$. Representing
$\overline{\phi}$ and $\phi$ as
\beqa
{\overline \phi} & \rightarrow &
\sigma_+ \theta \overline{\theta} \equiv
\left(
\begin{array}{ll}
0 & 0 \\
1 & 0
\end{array}
\right) \, \theta \overline{\theta} \, , \nonu
\phi & \rightarrow &
\sigma_- \theta \overline{\theta} \equiv
\left(
\begin{array}{ll}
0 & 1 \\
0 & 0
\end{array}
\right) \, \theta \overline{\theta} \, ,
\eeqa
we have $[\overline{\phi}\, , \, \phi] = - \sigma_3
\theta \overline{\theta} = - \left(
\begin{array}{lr}
1 & 0 \\
0 & -1
\end{array}
\right) \,
\theta \overline{\theta}$ and the Hamiltonian is diagonalized.
Denoting the 2-component Pauli spinor $\Phi$ in this basis:
\beq
\Phi \equiv
\Psi_{up} \left(
\begin{array}{l}
1 \\
0 \end{array}
\right)
+
\Psi_{down} \left(
\begin{array}{l}
0 \\
1 \end{array}
\right) \, .
\eeq
the ground state $\Psi_0$ is polarized: $\Psi_0 = \Psi_{up}
\left(
\begin{array}{l}
1 \\
0 \end{array}
\right)$. For $\Psi_0$, the ground state energy is zero.
To get an analytic form of $\Psi_{up}$, one considers the following
ground state equation:
\beq
{\cal X} \Psi_{up} \equiv
\left[ \dell{\theta} + \theta( \dell{z} + \frac{\overline{z}}{4} )
\right ] \Psi_{up} = 0,
\label{eq: ground}
\eeq
Notice that the ground state $\Psi_{up}$
is ``chiral" with
respect to the Grassmannian axes in the sense that
\beq
X \, \psi_{up} = \int d\theta \, {\cal X} \Psi_{up}.
\eeq
where $X = \dell{z} + \frac{\overline{z}}{4}$ as it should, and hence
$\psi_{up} = \Psi_{up}$ in this case.
One readily finds that $\Psi_{up} =  e^{-\frac{|z|^2}{4}}$
satisfies (\ref{eq: ground}) , for $\dell{\theta} \Psi_{up} = 0$,
$\theta ( \dell{z} + \frac{\overline{z}}{4} )
\, \Psi_{up} = 0$.
The result agrees with the standard {\em supersymmetric} quantum
mechanics of a particle in the superpotential
$W_z = - i \frac{\overline z}{2},
W_{\overline z} = i \frac{z}{2}$:
\beqa
Q \is ( \sqrt 2 P_z + \frac{1}{\sqrt 2} W_z ) \sigma_+ \, ,  \\
{\overline Q} \is ( \sqrt 2 P_{\overline z}
+ \frac{1}{\sqrt 2} W_{\overline z} ) \sigma_- \, , \\
H \is Q {\overline Q} + {\overline Q} Q \, .
\eeqa

Indeed, our spinning formalism is a variation of the same theme. The
only difference is that it allows us to find some non-trivial
ground states before integrating out the Grassmannian axes, as
will be seen in the case of spinning fractional quantum Hall effect.

\newsubsection{Spinning Representation}
\par
With this formulation, one can proceed to generalize the braid group
representation discussed in section 2. As will be explicit from the
wavefunctions to be calculated later on in section 5.2,
this generalization, though straightforward, is
non-trivial because the spin degree of freedom is incorporated.

For a start, let us consider two spinning
particles moving freely in a super-plane.
The winding number for a path going about a point $(z_0, \, \theta_0)$
in the super-plane is
\beq
\fr{2 \pi i} \, \int dz \, \int d \theta \,
\frac{\theta - \theta_0}{ z - z_0 - \theta \theta_0} \, .
\label{Superwind}
\eeq
Notice that $z-z_0-\theta\theta_0$ is even and $\theta - \theta_0$ odd.
They are respectively the even and odd intervals of the superplane.
Following \cite{SWZW}, denote a point in the superplane as
$Z$, we can formally write the super intervals as
\beqa
Z - Z_0 \, & \equiv & \, z - z_0 - \theta\theta_0 \, \nonu
( Z - Z_0 )^{\half} \, & \equiv & \, \theta - \theta_0 \, ,
\eeqa
which can be conveniently expressed in the following way:
\beq
( Z - Z_0 )^k = \left\{ \begin{array}{ll}
( z - z_0 - \theta\theta_0 )^k, & \hspace{2mm} \mbox{$k \in$ \bf Z} \\
(\theta - \theta_0 )( z - z_0 - \theta\theta_0 )^{k-\half},
& \mbox{ $k \in$ \bf Z + $\half$}.
                    \end{array}
              \right.
\eeq
In this notation,
the integrand $\frac{\theta - \theta_0}{z - z_0 - \theta\theta_0}$
can be seen as $(Z - Z_0 )^{- \half}$, and
(\ref{Superwind}) is formally
$\fr{2 \pi i} \int dZ (Z - Z_0)^{-\half}$ which looks more like the
expression for the usual winding number integral
$\fr{2 \pi i} \int dz (z - z_0)^{-1}$.
The ``reason" for $(Z - Z_0)^{-\half}$ instead of $(Z - Z_0)^{-1}$
is that $\int d Z $ is odd and we need an odd integrand to make the
whole integral even.
In our setup, it may be rewritten as
\beq
\fr{2 \pi i} \, \int d \overline{\theta} \, ( \overline{\theta} d \phi ) \,
\int d \theta \,
\frac{\theta - \theta_0}{ z - z_0 - \theta \theta_0} \, .
\eeq

With time $t$ as the parameter for the path, the spinning analogue
of the {\em charged}
winding number is
\beq
\fr{2 \pi i} \, \int_{t_0}^{t_1}  dt
\int d \overline{\theta} \,
\int d \theta \,
\frac{\theta - \theta_0}{ z - z_0 - \theta \theta_0} \,
( \overline{\theta} \dot{\phi} ) \, T \otimes T_0 \, ,
\eeq
and following the same procedure, we arrive at the spinning Lagrangian:
\beq
{\cal L} =
\half (\theta \dot{\overline{\phi}}) (\overline{\theta} \dot\phi)
+ {\cal A}_z (\overline{\theta} \dot\phi )
+ ( \theta \dot{\overline{\phi}} ) \overline{{\cal A}}_{\overline{z}}
+ {i\over{2}} (\dot\phi \overline\phi - \dot{\overline{\phi}} \phi),
\eeq
where, $l, \overline{l}$ being any real numbers,
\beqa
{\cal A}_z \is i l \frac{\theta - \theta_0}{z - z_0 - \theta \theta_0}
\, T \otimes T_0 \, , \nonu
\overline{{\cal A}}_{\overline z} \is i {\overline l} \frac{\overline{\theta}
- \overline{\theta}_0}{ \overline{z} - \overline{z}_0
- \overline{\theta} \overline{\theta}_0}
\, \overline{T} \otimes \overline{T}_0 \,.
\eeqa
As before, $T$ and $T_0$ are the respective
representations carried by the winding particle and its
counterpart which appears as a puncture.
In the Schr\"odinger representation, the spinning
Hamiltonian becomes
\beq
{\cal H} = \overline{\Pi} \, \Pi
- \Pi \, \overline{\Pi} \, ,
\eeq
where
\beqa
\Pi \is {\dell{\theta}} + \theta {\dell{z}} +  i {\cal A}_z, \nonu
\overline{\Pi} \is \dell{\overline{\theta}}
+ \overline{\theta} \dell{\overline{z}}
+ i \overline{\cal A}_{\overline{z}}.
\eeqa
We can readily
write down the Hamiltonian of $N$ spinning particles.
The zero-energy states of the $N$-body Hamiltonian
can be easily found from the first order
equations which are the supersymmetric generalization of the ones that
appeared in \cite{KZ}\cite{Lai-Ting}:
\beqa
\left[ \dell{\theta_i} + \theta_i \dell{z_i}
- l \sum_{ j = 1, \, j \neq i}^N
\frac{\theta_i - \theta_j}{z_i - z_j - \theta_i \theta_j}
\, T_i \otimes T_j \, \right] \Psi \is 0, \nonu
\left[ \dell{\overline{\theta_i}} + \overline{\theta_i} \dell{\overline{z_i}}
-{\overline l} \sum_{ j = 1, \, j \neq i}^N
\frac{\overline{\theta_i}
- \overline{\theta}_j}{ \overline{z_i} - \overline{z}_j
- \overline{\theta_i} \overline{\theta}_j}
\, \overline{T}_i \otimes \overline{T}_j \, \right] \Psi \is 0.
\label{super KZ}
\eeqa
These supersymmetric Knizhnik-Zamolodchikov equations have been discussed
extensively in the literature \cite{Fuchs}. They originate from the
null vectors of the combined representation of Kac-Moody algebra
and super Virasoro algebra.
In the path integral approach, it is explicit that they
give the covariant horizontality condition with respect to the
flat connection
\beq
\Omega = - l \sum_{ k = 1, \, k \neq j}^N
\frac{\theta_j - \theta_k}{z_j - z_k - \theta_j \theta_k}
\, T_i \otimes T_j \, dz_j  d\theta_j,
\eeq
which is the spinning analogue of the Kohno connection \cite{Kohno} of
a holomorphic bundle. We have therefore constructed a representation
of Artin's braid group $B_N$ with Feynman kernels of spinning particles.
The threads of $B_N$ correspond to the {\em spinning} world lines.
As in the spinless case, the representation space contains the
space of correlation functions of super WZW theories.
We have thus made an explicit link between spinning anyons and super WZW
model. The factorizable ground states of spinning anyons are given by the
exact solutions of super
Knizhnik-Zamolodchikov equations (\ref{super KZ}).

\newsection{Polarized Ground States of FQHE}
\fpar
The quantum Hall effect \cite{FQHE}\footnote{For a quick review of
quantum Hall effect, see appendix B of \cite{BEMS}.}
is a rather
unusual collective transport phenomenon of two-dimensional electron gas.
When the external
magnetic field is strong, the thermal fluctuation is suppressed at
low temperature, and the mobility of the charge carriers is high {\it etc.},
the Hall conductance has a staircase dependence on the
magnetic field strength. Concomitantly, the longitudinal
conductivity is practically zero at the centre of the plateau.
To understand the peculiarity of the Hall effect at these extreme
conditions, it is essential to find the
many-body ground state of the quantum system.
In the case of the integer quantum Hall effect,
the system is a collection of
simple harmonic oscillators. The
Landau level provides the necessary energy gap that
supports the plateaux of Hall conductivities at
integral multiples of ${e^2 \over \hbar}$.
However, for the fractional Quantum Hall effect (FQHE),
the incompressibility of the
liquid is less straightforward.
Additional ideas are needed to account for the
experimental discoveries of FQHE.

\newsubsection{Laughlin Ground State}
\par
The starting point of a plausible theory of FQHE is
Laughlin's ans\"atz \cite{Laughlin}:
\beq
|m \rangle
= \prod_{j<k} ( z_j - z_k )^m \exp (- \frac{1}{4 l^2} \sum_{i} | z_i |^2),
\label{L state}
\eeq
where $l = \sqrt{ \frac{ \hbar c }{e B}}$
is the magnetic length, $\hbar$, $c$ being the usual universal constants,
$e$ is the charge of the electron and $B$ is the strength of the magnetic
field. It is postulated that $|m \rangle$ is the ground state of the
electrons exhibiting FQHE with fractional filling factor $1 \over m$.
The reason why $m$ is odd is because $|m \rangle$ describes a system
of electrons which have fermionic statistics. In \cite{Ting-Lai},
we have proposed a Hamiltonian $H$ (\ref{Ham})
for which $|m \rangle$ is the
{\em exact} ground state. The $N$-body Hamiltonian contains
Kohno connection \cite{Kohno}
\beqa
A_{z_{j}} \is i m \hbar \sum_{k=1, k \ne j}^N
\frac{ T_{j} \otimes T_{k}}{ z_{j} - z_{k}}, \nonu
A_{\overline{z}_{j}} \is i \overline{m} \hbar \sum_{k=1, k \ne j}^N
\frac{ \overline{T}_{j} \otimes \overline{T}_{k}}{{\overline{z}_{j}} -
{\overline{z}_{k}}},
\eeqa
which reflects the topological properties of the
configuration space as mentioned in section 2.
Let $m^*$ be the effective mass of the electron,
$B_{z_j}, B_{{\overline z}_j}$ the components of the gauge field of
the external magnetic field, the Hamiltonian is
\beqa
\lefteqn{  H =
\frac{1}{m^{*}} \sum_{j=1}^N
[~( - i \hbar \partial_{z_{j}} + \frac{e}{c} B_{z_{j}} + A_{z_{j}} )
( - i \hbar \partial_{\overline{z}_{j}} + \frac{e}{c} B_{\overline{z}_{j}}
+ A_{\overline{z}_{j}}) } \nonu
& & \mbox{}
+ ( - i \hbar \partial_{\overline{z}_{j}} + \frac{e}{c} B_{\overline{z}_{j}}
+ A_{\overline{z}_{j}} )
( - i \hbar \partial_{z_{j}} + \frac{e}{c} B_{z_{j}}
+ A_{z_{j}} )~] \nonu
& =  &
\frac{\hbar^2}{m^{*}} \sum_{j=1}^N \left( D_{z_j} D_{\overline{z}_{j}} +
D_{\overline{z}_{j}} D_{z_j} \right) \, ,
\label{Ham}
\eeqa
where
\beqa
D_{z_j} & \equiv &
\partial_{z_j} + i \frac{e}{\hbar c} B_{z_j}
+ \frac{i}{\hbar} A_{z_j} \, , \\
D_{\overline{z}_{j}} & \equiv &
- \partial_{\overline{z}_{j}}
- i \frac{e}{\hbar c} B_{\overline{z}_{j}}
- \frac{i}{\hbar} A_{\overline{z}_{j}},
\eeqa
Now, since all the particles are indistinguishable, they
carry the same representation. Thus, for any two
particles $k$, $j$, we have $T_j  = T_k$ and
$T_j = \overline{T}_j$, $j = 1, \cdots, N$. One may use hermitian
matrices
to represent $T_j^\alpha$, $\alpha = 1, \cdots, \mbox{\rm dim} \, G$.
In the symmetric gauge,
\beqa
B_{z_{j}} \is - i \frac{B}{4} \overline{z}_{j}\, , \nonu
B_{\overline{z}_{j}} \is  i \frac{B}{4} z_{j}\, , \nonu
m \is - \overline{m}\, ,
\label{eq: symmetric gauge}
\eeqa
one calculates
the commutator of $D_{z_i}$ and $D_{\overline{z}_{j}}$:
\beq
\left[ D_{z_j}, D_{\overline{z}_{j}} \right] = \frac{eB}{2 \hbar c}
       + 2 \pi m
       \sum_{ k = 1, k \neq j}^N \, \delta^{(2)} ( z_j - z_k )
       T_j \otimes T_k.
\eeq
The term $\frac{eB}{2 \hbar c}$ is related to the
zero-point energy of a simple harmonic oscillator, whereas
the Dirac delta functions arise from the
2-dimensional Green function of the plane:
\beqa
\partial_{\overline{z}} \frac{1}{ z - w} \is - \pi \delta^{(2)}
( z - w )\, , \\
\partial_z \frac{1}{ {\overline z} - {\overline w}}
\is - \pi \delta^{(2)} ( z - w )\, .
\eeqa
With $\omega \equiv \frac{e B}{m^* c}$, we can rewrite (\ref{Ham}) as
\beq
H =  \frac{2 \hbar^2}{m^{*}}
\sum_j D_{\overline{z}_{j}} D_{z_j} + {N \over 2} \hbar \omega
+ \frac{ 2 \hbar^2}{m^{*}}
\pi m \sum_j \sum_{k=1, k \neq j}
\delta^{(2)} ( z_j - z_k ) T_j \otimes T_k.
\label{Ham-delta}
\eeq
Since this Hamiltonian is derived from the assumption that the
underlying configuration space is not simply connected,
the ground state of $H$ can be obtained by letting
$z_j \neq z_k$ for all $j$ and $k$, and then consider
the following first order equation for $j$-th electron:
\beq
D_{z_j} \psi_{0 \, j} = \left[ \partial_{z_j}
+ \frac{e B}{4 \hbar c} {\overline{z}_{j}}
- m \sum_{k=1, k \neq j}^N
\frac{T_j \otimes T_k}{z_j - z_k} \right]
\psi_{0 \, j}= 0.
\label{eq: GSE}
\eeq
Physical considerations require $f_j$ to be holomorphic.
As discussed by Laughlin \cite{Laughlin},
the many-body wavefunction comprises only
of single-body wavefunctions lying in the lowest Landau level.
This idealization is valid, in view of the facts that
there are only enough electrons to fill the lowest Landau level
and that the cyclotron energy $\hbar \omega$ is much greater than
Coulomb interaction. Overlaps with contributions from higher Landau levels
are practically negligible.
Writing
\beq
\psi_{0 \, j} =
       \exp ( - \frac{1}{4 l^2} |z_j|^2 )
       f_j(z_1, \cdots, z_N ) \, ,
\eeq
equation (\ref{eq: GSE}) then becomes
\beq
\partial_{z_j} f_j(z_1, \cdots, z_N )
- m \sum_{k=1, k \neq j}^N \frac{T_j \otimes T_k}{z_j - z_k}
f_j(z_1, \cdots, z_N ) = 0.
\label{eq: chiral KZ}
\eeq
Thus, we see that {\em chiral}
Knizknik-Zamolodchikov equations are relevant in FQHE.
(These equations
have also been used to explore the possibility
of non-abelian Aharanov-B\"ohm effect \cite{Verlinde}.)
For $T_j = 1, \, j = 1, \cdots N$,
the holomorphic function satisfying (\ref{eq: chiral KZ}) is
\beq
f_{j}(z_1, \cdots, z_N) = \mbox{\rm const}
\prod_{k = 1, k \neq j} ( z_j - z_k )^{m} \,.
\eeq
For $m > 0$, $f_j$ vanishes whenever $z_j$ coincides with any other $z_k$.
In other words, particle $j$ is kept apart from the other
electrons. This solution is consistent with the repulsive delta-function
potential $\sum_{ k = 1, k \neq j} \delta^{(2)} ( z_j - z_k )$,
because for any $j$,
\beq
\int d z_j d \overline{z}_j
\left(
\sum_{ k = 1, k \neq j} \!
\delta^{(2)} ( z_j - z_k )
\right) \, |f_j|^2 = 0.
\eeq
Though $f_j$ is not normalizable, $\psi_{0 \, j}$ is, thanks to
the factor $\exp ( - \frac{1}{4 l^2} |z_j|^2 )$ contributed by
the strong magnetic field. Solving $D_{z_j} \psi_{0 \, j} = 0$ for
arbitrary $j$, we find that the solution is exactly
the Laughlin wavefunction:
\beq
\psi_{0} = \mbox{\rm const.}
\prod_{j<k} ( z_j - z_k )^{m}
\exp (- \frac{1}{4 l^2} \sum_{i} | z_i |^2) \, .
\eeq

Because $\psi_0$ is the many-body wavefunction of electrons,
$m$ is an odd number.
{}From these results, one is able to identify the physical origin
of FQHE with filling fractions ${1 \over p}$, $p = m$:
Since the configuration space is multiply-connected, one has
to consider the minimal coupling of the
Kohno connection in addition to the electromagnetic gauge potential.
The factor $\prod_{j < k} (z_j - z_k)^{m}$ bears testimony to
the non-simply connected nature of the topology; Kohno connection
arises as homotopical labels of the paths in terms of
charged winding numbers \cite{Ting-Lai}.

\newsubsection{Spinning Analogue of the Laughlin State}
\par
While FQHE with odd $p$ stems from the braid group representation
associated with the non-simply connected configuation space $M_N$,
it is of interest to examine if the {\em spinning}
braid group representation
associated to particles with spin in the ``puncture" phase
will also yield FQHE. Put differently,
when the spin degree of freedom is turned on,
we want to know if there is an incompressible ground state
exhibiting FQHE. For this purpose, we consider
the Hamiltonian:
\beqa
H \is \sum_j \int d \overline{\theta}_j
d \theta_j \, {\cal H}_j \, , \\
\label{spin Ham}
{\cal H}_j \is \frac{\hbar^2}{m^*} \left(
\overline{\cal D}_j {\cal D}_j
- {\cal D}_j \overline{\cal D}_j
\right)
- g \mu B \, \sigma_3 \theta_j \overline{\theta}_j  \, .
\eeqa
Each ${\cal H}_j$ is the spinning Hamiltonian of particle $j$.
$g \mu B$ is the Zeeman energy, $g$ the $g$-factor and
$\mu$ denotes the magnetic moment of the electron.
In the symmetric gauge
(\ref{eq: symmetric gauge}), the covariant derivatives are
\beqa
{\cal D}_j \is
\dell{\theta_j} + \theta_j \left(
\dell{z_j} + \frac{eB}{4 \hbar c} {\overline z}_j \right)
- m \sum_{ k = 1, \, k \neq j}^N
\frac{\theta_j - \theta_k}{z_j - z_k - \theta_j \theta_k}
\, T_j \otimes T_k \, , \\
{\overline {\cal D}}_j \is
\dell{{\overline \theta}_j} + \overline{\theta}_j
\left(
\dell{{\overline z}_j} - \frac{eB}{4 \hbar c} z_j \right)
- \overline{m} \sum_{ k = 1, \, k \neq j}^N
\frac{\overline{\theta}_j - \overline{\theta}_k}
{ \overline{z}_j - \overline{z}_k - \overline{\theta}_j \overline{\theta}_k}
\overline{T}_j \otimes \overline{T}_k \, .
\eeqa
Now, write $D_j \equiv \dell{\theta_j} + \theta_j \dell{z_j}$,
$\overline{D}_j \equiv \dell{\overline{\theta}_j} +
\overline{\theta}_j \dell{\overline{z}_j}$, we have
$
D_j \overline{D}_j = - \overline{D}_j D_j
$, because $D_j$ and $\overline{D}_j$ are odd
differential operators.
With this consideration, the Green functions of the super-plane are
\beqa
D_j
\left(
\frac{1}{\overline{z}_j - \overline{z}_0
- \overline{\theta}_j \overline{\theta}_0 }
\right)
\is - \pi \delta^{2} ( z_j - z_0 - \theta_j \theta_0 ) \, , \\
\overline{D}_j
\left(
\frac{1}{z_j - z_0 - \theta_j \theta_0 }
\right)
\is + \pi \delta^{2} ( z_j - z_0 - \theta_j \theta_0 )  \, .
\eeqa
So in the symmetric gauge (\ref{eq: symmetric gauge}), and when all
the particles are identical, the anticommutator is
\beq
\{ {\cal D}_j \, , \, \overline{\cal D}_j \}
= - \frac{ e B}{ 2 \hbar c} \theta_j \overline{\theta}_j
- 2 \pi m \sum_{j = 1, k \neq j}
\delta^{2} ( z_j - z_k - \theta_j \theta_k )
T_j \otimes T_k \, .
\eeq
Again, we see that
Dirac delta functions appear. They prevent two
particles from occupying the
same point at the same instance in the super-plane.
The spinning Hamiltonian of particle $j$ becomes
\beqa
{\cal H}_j \! & = & \!
\frac{2 \hbar^2}{m^*}
\overline{\cal D}_j {\cal D}_j
+ \half \hbar \omega \theta_j \overline{\theta}_j
- g \mu B \, \sigma_3 \theta_j \overline{\theta}_j \nonu
&\quad &
+ \frac{2 \hbar^2}{m^*} \pi m \! \sum_{j = 1, k \neq j}
\delta^{2} ( z_j - z_k - \theta_j \theta_k )
T_j \otimes T_k \, .
\eeqa
It is implicit in the Hamiltonian that the spin of each electron
is aligned either parallel (up) or
anti-parallel (down) with respect to the external magnetic field.
Only the spin components normal to the direction of
the magnetic field enter as
dynamical variables. To find the spin-polarized ground state with
zero energy, we need to consider $\Psi_{up}$
such that for arbitrary $j$, ${\cal D}_j \Psi_{up} = 0$.
In addition, due to the presence of the repulsive interaction
term of infinitesimal range,
$\Psi_{up}$ must contain a factor
which is some positive power of $(z_j - z_k - \theta_j \theta_k)$.
As before, we write $T_j = 1, j = 1, \cdots, N$, and
\beq
{\cal F}_j = \mbox{\rm const.} (-1)^{j-1}
\prod_{1 \leq j < k \leq N}
( z_j - z_k - \theta_j \theta_k )^m \, .
\eeq
It is easy to show that
\beq
\Psi_j^{up} = \int
d\theta_1 \cdots d\theta_{j-1} d\theta_{j+1}
\cdots d \theta_N \, \overline{\theta}_j
\, \exp (-\frac{1}{4 l^2} \sum_i |z_i|^2 )
{\cal F}_j
\eeq
satisfies the ground state equation:
\beq
{\cal D}_j \Psi_j^{up}
= \left[ \dell{\theta_j} + \theta_j \left(
\dell{z_j} + \frac{eB}{4 \hbar c} {\overline z}_j \right)
- m  \sum_{ k = 1, \, k \neq j}^N
\frac{\theta_j - \theta_k}{z_j - z_k - \theta_j \theta_k}
\, \right] \Psi_j^{up} = 0.
\eeq
In particular, it is worth remarking that ${\cal F}_j$ is the conformal
block of the super $U(1)$ current algebra:
\beq
\left(
D_j  - m  \sum_{ k = 1, \, k \neq j}^N
\frac{\theta_j - \theta_k}{z_j - z_k - \theta_j \theta_k}
\, \right) {\cal F}_j = 0 \, .
\eeq
Using the many-body analogue of (\ref{eq: spin eigen}), (\ref{psi}), namely
\beq
H \psi_0^{up} = \sum_j \left( \int d\overline{\theta}_j d\theta_j
{\cal H}_j \Psi_j^{up} \right) \, ,
\eeq
we have
\beq
\psi_0^{up} = N \mbox{const.} \int \prod_{j=1}^{N} d \theta_j
\prod_{1 \leq j < k \leq N} (z_j - z_k - \theta_j \theta_k)^m
\, \exp (-\frac{1}{4 l^2} \sum_i |z_i|^2 ) \, .
\label{MR state}
\eeq
Now, since $\theta_j^2 = \overline{\theta}_j^2 = 0$ for all $j$,
we see that
\beqa
\prod_{1 \leq j < k \leq N} (z_j - z_k - \theta_j \theta_k)^m
\is
\prod_{1 \leq j < k \leq N} \left[
(z_j - z_k )^m \left( 1 - \frac{\theta_j \theta_k}
{z_j -z_k} \right)^m \right] \nonu
\is
\prod_{1 \leq j < k \leq N} (z_j - z_k )^m
\prod_{1 \leq j < k \leq N} \left( 1  - m \frac{\theta_j \theta_k}
{z_j -z_k} \right) \, .
\label{eq: spin factor}
\eeqa
In the expansion of (\ref{eq: spin factor}),
the terms that do not vanish under the operation
$\int \prod_{j=1}^N d\theta_j$ must contain
$\prod_{j=1}^N \theta_{\sigma (j)}$. This
is possible only if $N$ is an even number.
In this case,
\beqa
\int d \theta_1 \cdots d \theta_N
\prod_{1 \leq j < k \leq N} \left( 1 - m \frac{\theta_j \theta_k}
{z_j -z_k} \right)
\is { m^N \over { 2^{N \over 2} ({ N \over 2} !) }}
\sum_{\sigma} (-1)^{\sigma} \left( \frac{1}{z_{\sigma_1} -
z_{\sigma_2}} \right) \cdots
\left( \frac{1}{z_{\sigma_{N-1}} - z_{\sigma_N}} \right) \nonu
& \equiv & m^N \, \mbox{\rm Pf} \, \left( \frac{1}{z_j - z_k} \right)
\eeqa
Here $\sigma$ runs over permutations of the $N$ indices,
$(-1)^{\sigma}$ is the parity of the permutation.
The expression Pf$(M_{jk})$ is called the Pfaffian of
an antisymmetric $N \times N$ matrix $M$ with entries $M_{jk}$.
Now, $\psi_0^{up} ( z_1, \cdots, z_N)$ is the physical wavefunction
describing an ensemble of electrons. Any interchange of arbitrary
pair of coordinates must result in a negative sign as Pauli principle
says. Consequently, $m$ must be an even number since
Pfaffian is antisymmetric.
We remark that (\ref{MR state}) is exactly the same as
the Moore-Read ans\"atz for spin polarized FQHE states at
even denominator filling fractions.

In this manner,
we have unveiled the physical origin of the Laughlin states\cite{Ting-Lai}
and the Moore-Read
states. The non-trivial topology of the configuration
space of $N$ electrons in the ``puncture" phase is manifested in the
Hamiltonians
(\ref{Ham-delta}) and (\ref{spin Ham}). Respectively, they
yield the Laughlin state and the Moore-Read state as exact non-degenerate
ground state solutions.

\newsection{Topological Excitations\footnote{Throughout the paper, we
only mention quasi-hole excitation. The quasi-particle is taken to
be the particle-hole conjugate of the quasi-hole.}}
\fpar
One of the necessary conditions for a many-body ground state
to display FQHE is that its quantum excitations are
massive. Among other things, it behooves the system to
be non-degenerate across a sufficiently finite range of variation
in the background magnetic field strength. In other words,
the collection of electrons in the ``puncture" phase must be capable
of buffering a certain amount of excess or deficiency in the quantum
flux tubes in the form of excited states in the energy spectrum. The crucial
point is that these excited states must lie within the large gap of
$\hbar \omega$ between two neighbouring Landau levels, if FQHE plateaux
are to take shape. The existence of such a substratum structure
superimposed over the Landau levels of a collection of oscillators
is a key to the understanding of FQHE.

In \cite{Laughlin}, Laughlin gave an ans\"atz of the wavefunction which
is a 1-quasi-hole excitation of the ground state $|m \rangle$ (\ref{L state}):
\beq
\psi_m (w; z_1, \cdots , z_N) = \prod_{j = 1}^N ( z_j - w )
|m \rangle \, ,
\eeq
where $w$ is the position of the quasi-hole. The existence of the quasi-hole
excitation is demonstrated in the {\em gedanken} experiment.
An infinitesimally thin solenoid is pierced through the ground state
$|m \rangle$ at position $w$. Adiabatically, a flux quantum $\frac{hc}{e}$
is added; $|m \rangle$ evolves in such a way that it remains an
eigenstate of the changing Hamiltonian. After the flux tube is
completely installed, the resulting Hamiltonian is related
to the initial one by a (singular) gauge transformation. To get
back to the original Hamiltonian, the
flux tube is gauged away, leaving behind an excited state
$\psi_m (w; z_1, \cdots, z_N)$. This idea is strongly
reminiscent of the Aharanov-B\"ohm effect.

The interesting and strange feature of FQHE is that
the charge $q_h$ of the quasi-hole is fractional.
The exact value can be determined via the plasma analogy. The
square of the wavefunction $\psi_m$ can be interpreted as
a probability distribution function of a plasma:
\beq
| \psi_m (w; z_1, \cdots, z_N) |^2 = e^{- \beta E},
\eeq
where $\beta = m$ plays the role of inverse temperature, and
the Gibbs energy $E ( w; z_1, \cdots, z_N)$
is given by
\beqa
E ( w; z_1, \cdots, z_N) & = & - 2 \sum_{j < k} \log | z_j - z_k |
+ \frac{1}{2 m l^2} \sum_j |z_j|^2 \nonu
& \quad &
- \frac{2}{m} \sum_{i=1}^N \log | z_i - w | \, .
\label{eq: plasma energy}
\eeqa
This is the total energy of a gas of $N$ classical
particles each carrying charge $q=1$ plus a
particle of charge $q_h = \frac{1}{m}$ which repel
each other through the 2-dimensional
``Coulomb" potential $- 2 \sum_{j < k} \log | z_j - z_k |$
in a uniform neutralizing background charge of density
$\rho_0 = \frac{1}{2 \pi m l^2}$.
It is clear that the first two terms in
(\ref{eq: plasma energy}) are contributed by the ground state $|m \rangle$.
The charge $q$ being 1 is related to the fact that representation
$T_j = 1$ is chosen for each electron.

Motivated by this physical picture, we can carry the plasma analogy
further and consider the same Hamiltonian (\ref{Ham}) for the
1-quasi-hole excitation but with a gauge transformed $A_{z_j},
A_{\overline{z}_j}$:
\beqa
A_{z_{j}} & \rightarrow & i m \hbar \sum_{k=1, k \ne j}^N
\left(
\frac{1}{ z_{j} - z_{k}}
+ \frac{\frac{1}{m}}{ z_{j} - w}
\right) \, ,
\nonu
A_{\overline{z}_{j}} & \rightarrow & -i m \hbar \sum_{k=1, k \ne j}^N
\left(
\frac{1}{{\overline{z}_{j}} - {\overline{z}_{k}}}
+ \frac{\frac{1}{m}}{{\overline{z}_{j}} - {\overline{w}}}
\right) \, .
\eeqa
It can be easily verified that $\psi_m$ satisfy the ground state
equations, $j = 1, \cdots, N$:
\beq
\left[ \partial_{z_j}
+ \frac{e B}{4 \hbar c} {\overline{z}_{j}}
- m \sum_{k=1, k \neq j}^N \left(
\frac{1}{z_j - z_k} + \frac{\frac{1}{m}}{z_j - w}
\right) \right]
\psi_m = 0.
\eeq
Remember that $A_{z_j}, A_{\overline{z}_j}$ comes from the
{\em charged} winding number constraint of
the paths of particle $j$ in the multiply-connected
configuration space. Attaching an additional solenoid on $|m \rangle$
therefore results in a new configuration space.
In other words,
electron $j$ sees the quasi-hole as a puncture as well, but
this time with charge $q_h = {1 \over m}$.
The excitation is topological in nature.
When a quasi-hole develops, the configuration space is topologically changed.
It is no longer $M_N$, but $M_{N+1}$.

So far, we are only concerned with
one quasi-hole excitation at $w$.
What about the wavefunctions of two or more quasi-holes?
As discussed by Halperin \cite{Halperin}, these multi-excitation states
should be an analytic function of the coordinates
of the electrons $z_1, \cdots, z_N$, and of the
quasi-holes $w_1, \cdots, w_{N_h}$ up to exponential
factors. The analytic condition is to require that
even the excitation wavefunctions should
only come from the lowest Landau level.
The Halperin ans\"atz is
\beq
\psi_m( w_1, \cdots, w_{N_h} )
= \prod_{1 \leq j<k \leq N_h}
( w_j - w_k )^{1 \over m}
\exp (- \frac{1}{4 m l^2} \sum_{i} | z_i |^2)
\prod_{j, k} ( w_j - z_k )
|m \rangle \, .
\eeq
If we write
\beq
|\frac{1}{m} \rangle =
\prod_{1 \leq j<k \leq N_h}
( w_j - w_k )^{1 \over m}
\exp (- \frac{1}{4 m l^2} \sum_{i} | w_i |^2) \, ,
\eeq
which is of the same form as Laughlin's ground state $| m \rangle$,
we find that
\beq
\psi_m = |\frac{1}{m} \rangle \, | m \rangle
\prod_{j, k} ( w_j - z_k ) \, .
\eeq
Written in this form, the physical content of a collection of
quasi-holes $|\frac{1}{m} \rangle $
is explicit: they are just ``electrons" of (representation) charge
$q_h = {1 \over m}$ each in the ``puncture" phase! The quasi-holes are also
under the influence of the external magnetic field, for the exponential
factor $\exp (- \frac{1}{4 m l^2} \sum_{i} | w_i |^2)$ is
required to make the wavefunction well defined under normalization.
The Hamiltonian for two species of electrons labelled by $q=1$
and $q_h = {1 \over m}$ is
\beqa
H \! &=& \!  \frac{ 2 \hbar^2}{m^{*}}
            \sum_j^N D_{\overline{z}_{j}} D_{z_j}
            + {N \over 2} \hbar \omega
            + \frac{ 2 \hbar^2}{m^{*}}
            \pi m \sum_j \sum_{k=1, k \neq j}
            \delta^{(2)} ( z_j - z_k )  \nonu
\! & \quad &\!
+ \frac{ 2 \hbar^2}{m_{h}}
\sum_j^{N_h} d_{\overline{w}_{j}} d_{w_j}
+ { N_h \over 2 } \hbar \omega_h
+ \frac{ 2 \hbar^2}{m_h}
\pi \frac{1}{m} \sum_j \sum_{k=1, k \neq j}
            \delta^{(2)} ( w_j - w_k )  \nonu
\! & \quad &\!
+  2 \hbar^2 \pi \left( { 1\over m^*} + {1 \over m_h} \right)
\sum_j^{N_h} \sum_k^N \delta^{(2)} ( w_j - z_k ) \, .
\label{eq: qhole}
\eeqa
where
\beqa
D_{z_j} \is
\partial_{z_j}
+ \frac{e B}{4 \hbar c} {\overline{z}_{j}}
- m \sum_{k=1, k \neq j}^N
\frac{1}{z_j - z_k}
- m \sum_{k=1}^{N_h}
\frac{\frac{1}{m}}{z_j - w_k} \nonu
d_{w_j} \is
\partial_{w_j}
+ \frac{\frac{e}{m} B}{4 \hbar c} {\overline{w}_{j}}
- m \sum_{k=1, k \neq j}^{N_h}
\frac{\frac{1}{m} \times \frac{1}{m}}{w_j - w_k}
- m \sum_{k=1}^{N}
\frac{\frac{1}{m}}{w_j - z_k} \, ,
\eeqa
with similar expressions for $D_{{\overline z}_j}$
and $d_{{\overline w}_j}$. We have denoted the ``mass" of a
quasi-hole as $m_h$, and $\omega_h = \frac{1}{m} \frac{e B}{ m_h c}$ is
the angular frequency of the cyclotron motion of the quasi-holes.

It is readily shown that $\psi_m ( w_1, \cdots , w_{N_h}\, ; \,
z_1, \cdots, z_N )$ is the {\em exact} ground state solution of
$H$ (\ref{eq: qhole}).
If these many-quasi-hole wavefunctions describe real physics as
Halperin suggested, so does $H$.
Furthermore, the path integral representation approach which
reveals the relevance and meaning of the Kohno connection allows one to
see explicitly that quasi-holes behave as if they were
spinless particles of charge $- \frac{e}{m}$ in the ``puncture" phase.
The picture which emerges from $H$ (\ref{eq: qhole}) may be captured
in Figure 6.
Our results show that
the topological excitation also has a Landau level structure
for its spectrum. The FQHE ground state thus comprises of
two species of ``punctures",
namely, electrons and quasi-holes.
The ground state energy of the quasi-hole
is the energy gap responsible for the incompressibility of the
FQHE liquid.
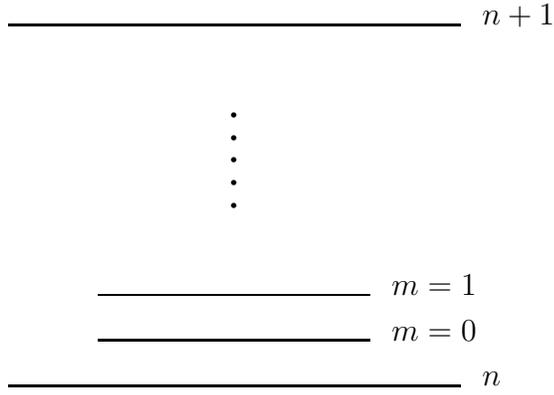
\gap{0}{Fig 6: The energy spectrum of FQHE. $n$ is
the partially filled Landau level of electrons.
$m$ labels the Landau level of quasi-holes.}

In order to support this interpretation, one has to have an answer to
the pressing question: What is the mass $m_h$ of the quasi-hole ?

As we learn in nuclear physics, the binding energy of nucleons can
be equated with $\delta m \, c^2$ if $\delta m$ is the
mass difference between a nucleus and the total of its fission
moities. It is tempting to apply this popularly known
$\delta E = \delta m \, c^2$
formula to FQHE as well:
\beq
\delta E = \hbar \omega_h = m_h \, c^2.
\eeq
The mass of the electron $m_e$ in the crystal lattice is
not its rest mass in the vacuum but gets modified to
$m^* = x m_e$ where $x$ is a dimensionless number.
By the same token, since the quasi-hole excitation is treated as if it is
some spinless electron with fractional charge, $m_h$
must also be modified to $m_h^* = y m_h$ for some empirical factor $y$.
Then, we find that the energy gap of the quasi-hole excitation is
\beq
\delta E = y \, C \sqrt B \, ,
\label{gap}
\eeq
with $C = \sqrt{\frac{e}{m} \hbar c }$. It is interesting to note
that this interpretation also leads to a square root dependence of
the energy gap $\delta E$ on $B$.
The proportional constant $C$ is determined solely by the
absolute value of the fractional charge $\frac{e}{m}$ and
the universal constants.

Except for the threshold\footnote{
A possible origin of threshold is discussed in section
7.2.}, the $\sqrt B$ dependence
is quite in line with experiments \cite{Stormer} \cite{FQHE}.
Of course, the many-body quantum mechanics we have here
is oversimplified in the sense that the imperfections of the GaAs-AlGaAs
heterostructure, the thickness of the heterojunction,
the mixing of higher Landau levels {\it etc} are neglected.
Nevertheless the main characteristics of
FQHE such as the quantum statistics, the
exact value of the fractional charge are sufficiently robust
even in the presence of those complications and
the plausibility of a simple Hamiltonian like
(\ref{eq: qhole}) is warranted.

One of the implications of expression (\ref{gap}) is
that the ratio of the energy gaps of $\nu_a = \frac{1}{m_a}$
and $\nu_b = \frac{1}{m_b}$ FQHE is given by
\beq
\sqrt{\frac{m_b}{m_a} \frac{B_a}{B_b}}
\label{ratio}
\eeq
where $m_a$ and $m_b$ are both odd numbers and
$B_a$ and $B_b$ are the magnetic field strengths at the centres of the
respective FQHE plateaux.

In an analogous fashion, we can also study the quasi-hole excitation
of the Moore-Read state. The spinning analogue of (\ref{eq: qhole})
is
\beqa
{\cal H} \! &=& \!
\sum_j^N \mbox{\Huge [}
\frac{ 2 \hbar^2}{m^{*}} {\cal D}_{\overline{z}_{j}} {\cal D}_{z_j}
+ \half \hbar \omega \theta_j \overline{\theta}_j
- g \mu B \, \sigma_3 \theta_j \overline{\theta}_j
+ \frac{ 2 \hbar^2}{m^{*}}
\pi m \sum_{k=1, k \neq j}
\delta^{(2)} ( z_j - z_k - \theta_j \theta_k )
\nonu
\! & \quad &\!
+  2 \hbar^2 \pi { 1\over m^*}
\sum_k^{N_h} \delta^{(2)} ( z_j - w_k - \theta_j\eta_k)
\mbox{\Huge ]}
\nonu
\! & \quad &\!
+ \sum_j^{N_h}
\mbox{\Huge [}
\frac{ 2 \hbar^2}{m_{h}} \Delta_{\overline{w}_{j}} \Delta_{w_j}
+ \half \hbar \omega_h \eta_j \overline{\eta}_j
- g_h \mu_h B \, \sigma_3 \eta_j \overline{\eta}_j
+ \frac{ 2 \hbar^2}{m_h}
\pi \frac{1}{m} \sum_{k=1, k \neq j}
\delta^{(2)} ( w_j - w_k - \eta_j \eta_k)
\nonu \! & \quad &\!
+  2 \hbar^2 \pi {1 \over m_h}
\sum_k^N \delta^{(2)} ( w_j - z_k - \eta_j\theta_k)
\mbox{\Huge ]} \, ,
\label{eq: sqhole}
\eeqa
where ${\cal D}_{z_j}$, $\Delta_{w_j}$ are the Grassmannian odd
covariant derivatives:
\beqa
D_{z_j} \is
\dell{\theta_j}
+ \theta_j \left(
\dell{z_j} +
\frac{e B}{4 \hbar c} {\overline{z}_{j}}
\right)
- m \sum_{k=1, k \neq j}^N
\frac{\theta_j - \theta_k}{z_j - z_k - \theta_j \theta_k}
- m \sum_{k=1}^{N_h}
\frac{\frac{1}{m}(\theta_j - \eta_k)}{z_j - w_k - \theta_j \eta_k} \nonu
\Delta_{w_j} \is
\dell{\theta_j}
+ \eta_j \left(
\dell{w_j}
+ \frac{\frac{e}{m} B}{4 \hbar c} {\overline{w}_{j}}
\right)
- m \sum_{k=1, k \neq j}^{N_h}
\frac{\frac{1}{m} \times \frac{1}{m}(\eta_j - \eta_k)}
{w_j - w_k -\eta_j \eta_k}
- m \sum_{k=1}^{N}
\frac{\frac{1}{m}(\eta_j - \theta_k)}{w_j - z_k - \eta_j \theta_k} \, .
\nonu
\eeqa
The many-quasi-hole wavefunction that is the ground state solution of
$\cal H$ (\ref{eq: sqhole}) is
\beqa
\! & & \!  (N + N_h) \mbox{\rm const.} \,
\int \prod_{j=1}^N d \theta_j \prod_{j=1}^{N_h} d \eta_j
\prod_{j < k} ( z_j - z_k - \theta_j \theta_k)^m
\prod_{j < k} ( w_j - w_k - \eta_j \eta_k)^{1 \over m}
\nonu \! & \quad &\!
\hspace{10mm} \times
\prod_{j, k} ( w_j - z_k - \eta_j \theta_k )
\exp ( - {1 \over {4 l^2}} \sum_i |z_i|^2
 - {1 \over {4 m l^2}} \sum_i |w_i|^2 )
\nonu \! & \quad &\!
= ( N + N_h)
m^{N - N_h} \,
\mbox{\rm const.} \,
\prod_{j < k} ( z_j - z_k)^m
\prod_{j < k} ( w_j - w_k)^{1 \over m}
\prod_{j \, , k} ( w_j - z_k)
\nonu \! & \quad &\!
\hspace{10mm} \times \exp \left( - {1 \over {4 l^2}} \sum_i |z_i|^2
 - {1 \over {4 m l^2}} \sum_i |w_i|^2 \right) \,
\mbox{\rm Pf} \, (M_{ij}) \,
\eeqa
where $M_{ij} = { 1 \over u_i - u_j }$, $u_i$ being
the combined set of $z_i$, $i = 1, \cdots, N$
and $w_i$, $i = 1, \cdots, N_h$. We emphasize that both
$N$ and $N_h$ must be even numbers. Therefore in the spinning case,
the quasi-hole excitations are paired.

\newsection{FQHE Ground State of Spin Singlets}
\fpar
So far, we are only concerned with spin-polarized ground states, \ie
all the spins align themselves parallel to the magnetic field normal
to the plane. In view of the large Zeeman energy when the magnetic
field is strong,
it is justifiable to assume that the spins are fully polarized.

However, experimental data reveal that
partially polarized FQHE ground states also exist. In particular, FQHE at a
{\em shared} filling
factor of $8 \over 5$ was observed to transit from a spin-unpolarized
state to a polarized one when the specimen was tilted
with respect to the magnetic field \cite{8/5}. This experimental
result is quite in line with Halperin's original suggestion
\cite{Halperin}: The $g$-factor of GaAs is rather small; hence,
when the magnetic field is not too strong, spin unpolarized states
should be viable. In this scenario, Zeeman energy cost
$g \mu B$ is low enough for some spins to get reversed.

To accommodate the spin degree of freedom parallel or anti-parallel
to the external field, it is useful to consider the Hamiltonian
(\ref{Ham}), or equivalently (\ref{Ham-delta}), with each $T_j$ carrying
the representation of $U(2)$ which is isomorphic to $SU(2) \times U(1)$.
As discussed earlier, we ignore the
Zeeman energy which is of the same order of
magnitude as the static Coulomb energy at characteristic length (magnetic
length $l$ ); for the time being, we just want to
study the {\em topological} ``interaction".
Intuitively, it is not hard to realize that
such Hamiltonian corresponds to the situation
where each electron carries a spin-$\half$ (highest weight)
representation of $SU(2)$ and
a $U(1)$ charge. In the ``puncture" phase, when one electron moves around
the other, a {\em non-abelian} charged winding number (\ref{windnum})
furnishes a topological label for the path; not only does an electron
see the $U(1)$ charges, it also perceives the spins on other electrons.

In the non-abelian analogue,
the ground state solution of such Hamiltonian is found by solving the
equation for all particles $j$:
\beq
\left[ \partial_{z_j}
+ \frac{e B}{4 \hbar c} {\overline{z}_{j}}
- \ell_{spin} \sum_{k=1, k \neq j}^N \frac{T_j \otimes T_k}{z_j - z_k}
- \ell_{charge} \sum_{k=1, k \neq j}^N \frac{1}{z_j - z_k}
\right] \psi_{0 \, j}= 0.
\label{eq: SGSE}
\eeq
{}From the physical viewpoint, the fundamental weights of the representations
have to be chosen in such a way that the resulting
wavefunction is a singlet. This is the non-abelian analogue of
the neutrality condition in the Coulomb gas picture of 2-dimensional
conformal field theory.
Using the Fierz identity for the Hermitian generators $T^{\alpha}$,
$\alpha = 1, \cdots, n^2 - 1$ in the
fundamental representation of $SU(n)$,
\beq
( T^{\alpha} )_{a}^{b} ( T^{\alpha} )_{c}^{d}
= \half \left( \delta_a^d \delta_c^b - {1 \over n} \delta_a^b \delta_c^d
\right) \, ,
\eeq
the ground state equation becomes
\beqa
\mbox{\Huge [}
\partial_{z_j}
+ \frac{e B}{4 \hbar c} {\overline{z}_{j}}
\hspace{-10mm} & &
- \,
\ell_{spin} \left(- \frac{n+1}{2n} \right) \sum_{k=1, k \neq j}^{N \over 2}
\frac{1}{z_j^{\uparrow} - z_k^{\uparrow}}
- \ell_{spin} \frac{n^2 - 1}{2n} \sum_{k=1}^{N \over 2}
\frac{1}{z_j^{\uparrow} - z_k^{\downarrow}}
\nonu \hspace{-10mm} & \quad &
- \,
\ell_{charge} \sum_{k = 1 \, , k \neq j}^{N \over 2}
\frac{1}{z_j^{\uparrow} - z_k^{\uparrow}}
-\ell_{charge} \sum_{k = 1}^{N \over 2}
\frac{1}{z_j^{\uparrow} - z_k^{\downarrow}}
\mbox{\Huge ]}
\psi_{0 \, j}= 0 \, ,
\label{eq: Halperin}
\eeqa
or
\beqa
\mbox{\Huge [}
\partial_{z_j}
+ \frac{e B}{4 \hbar c} {\overline{z}_{j}}
\hspace{-10mm} & &
- \ell_{spin} \frac{n^2 - 1}{2n} \sum_{k=1}^{N \over 2}
\frac{1}{z_j^{\downarrow} - z_k^{\uparrow}}
- \ell_{spin} \left(- \frac{n+1}{2n} \right) \sum_{k=1, k \neq j}^{N \over 2}
\frac{1}{z_j^{\downarrow} - z_k^{\downarrow}}
\nonu \hspace{-10mm} & \quad &
-\ell_{charge} \sum_{k = 1}^{N \over 2}
\frac{1}{z_j^{\downarrow} - z_k^{\uparrow}}
-\ell_{charge} \sum_{k = 1 \, , k \neq j}^{N \over 2}
\frac{1}{z_j^{\downarrow} - z_k^{\downarrow}}
\mbox{\Huge ]}
\psi_{0 \, j}= 0 \, ,
\label{eq: Halperin1}
\eeqa

Now, if we let $\ell_{spin} = - \frac{2}{n + k}$ with $k = 1$,
the spin portion of (\ref{eq: Halperin}) and
(\ref{eq: Halperin1}) can be identified with the
bosonization of free fermions carrying the representation of $SU(n)$.
With this choice and $\ell_{charge} = q + \half$,
the contribution of spin as $SU(2)$ in FQHE ground state combines
with the $U(1)$ winding number label as follows.
\beq
\left[
\partial_{z_j}
+ \frac{e B}{4 \hbar c} {\overline{z}_{j}}
-( \ell_{charge} + \half ) \sum_{k = 1 \, , k \neq j}^{N \over 2}
\frac{1}{z_j^{\uparrow} - z_k^{\uparrow}}
-( \ell_{charge} - \half) \sum_{k = 1}^{N \over 2}
\frac{1}{z_j^{\uparrow} - z_k^{\downarrow}}
\right]
\psi_{0 \, j}= 0 \, ,
\eeq
and a corresponding expression for (\ref{eq: Halperin1}).
Setting $\ell_{charge} = q + \half$, we find that
Halperin state \cite{Halperin} given by
\beqa
\psi_{mmn} (z_1^{\uparrow}, \cdots , z_{N \over 2}^{\uparrow} \, ;
z_1^{\downarrow}, \cdots, z_{N \over 2}^{\downarrow} )
\! & = & \! \prod_{j < k}^{N \over 2}
(z_j^{\uparrow} - z_k^{\uparrow} )^p
(z_j^{\downarrow} - z_k^{\downarrow} )^p
\prod_{r, s}^{N \over 2} ( z_r^{\uparrow} - z_s^{\downarrow} )^q
\nonu \! & \quad & \!
\times \exp \left( - \frac{1}{4 l^2} \sum_i^{ N \over 2}
|z_i^{\uparrow}|^2 + |z_i^{\downarrow}|^2 \right)
\eeqa
turns out to be the {\em exact}
ground state solution with $p$ constrained as
$p = q + 1$.
It is interesting to mention that the same constraint is
discussed by Girvin using the Fock cyclic condition
in the appendix of reference \cite{FQHE}.
Also, the Halperin state has
been constructed {\it a priori} in terms of the conformal block of $k=1$
$SU(2)$ WZW theory and that of the rational torus at level $2q + 1$ \cite{MR}.
We have shown that, starting from an appropriate Hamiltonian which describes
a system of electrons in the ``puncture" phase, there is no mystery
why a conformal field theory with \( \widehat{SU(2)}_{k=1} \) symmetry
can be used to produce the wavefunction of a non-relativistic
phenomenon.

Similar to what happened to the Laughlin's state,
the zero-range delta potential requires $q$ to be positive. The filling
fraction of Halperin state is $\frac{2}{2q + 1}$, with $q$ an even
number since electrons are fermions.
It is likely that the unpolarized FQHE state with
filling fraction $1 + \frac{3}{5}$ observed in the real world \cite{8/5}
is the particle-hole conjugate of Halperin state with $q = 2$.
Following the same line of thought of the previous section,
the quasi-hole excitation
of the Halperin state can be ascertained to
be characterized by a fractional (representation) charge of
$\frac{1}{2q+1}$ and spin $\half$.

\newsection{Discussions}

\newsubsection{Connection with WZW models}
\par
{}From the quantum mechanics of a system of $N$ particles in
the collective ``puncture" phase,
the zero-energy equations of (\ref{Sch}), namely (\ref{KZ1})
and (\ref{KZ2}) determine the {\em factorizable} $N$-body wavefunctions.
In this special case, the outcome is the same as the 3-dimensional
Chern-Simons guage theory \cite{Witten}\cite{CS-theory}. With a
suitable value chosen for $k$, solving the equation
(\ref{KZ1}) gives $\psi$ as the conformal blocks of the
WZW theory.
The quantum mechanics of $N$ punctures give yet another
3-dimensional description of 2-dimensional conformal field theories.
However, unlike the previous correspondence of Chern-Simons theory
with the {\em chiral} moiety of the WZW theory, $\psi$ has to
satisfy (\ref{KZ2}) as well. In addition, since a quantum mechanical
wavefunction must be invariant with respect to monodromy, $\psi$
may be identified with the correlation function of a WZW theory.
Analogously, the spinning version of the
path integral representation of the braid group admits the space of the
correlation functions of a super WZW theory as the representation
space.

In the representation theory of current algebra,
Knizknik-Zamolodchikov equations originate from the
existence of null vectors of the combined conformal and
Kac-Moody algebras \cite{KZ}.
Though these first-order differential equations are not sufficient
to determine the operator content of a WZW theory, they nevertheless
provide a way to calculate the $N$-point function
of the fields corresponding to the {\em integrable} representation
of the theory \cite{Gepner-Witten}. The correlation
function of a non-integrable field with any other fields vanishes,
indicative of a selection rule in the theory. It follows that the
Knizknik-Zamolodchikov equations supplemented with a set of
algebraic equations yield a solution space which is identical
to the Hilbert space of the WZW theory \cite{Gepner-Witten}
\cite{Tsuchiya-Kanie}.

Now, when the group manifold $G$ is $U(1)$, WZW theory reduces
to a boson compactified on a circle.
In this case, the conformal field theory is a representation
of a chiral algebra called rational torus or $U(1)$ current algebra.
The null vectors
of the {\em purely} Kac-Moody algebra do not tell much
story except that the correlation functions must be singlets.
Therefore, for $U(1)$ charges,
Knizknik-Zamolodchikov equations are sufficient
to determine the operator content of the corresponding WZW theory
with central charge $c = 1$.

Because of this connection, we can understand why it is possible to use
the conformal blocks of rational torus \cite{MR} or
the vertex operators of string theory \cite{Fubini}
to construct the Laughlin wavefunctions. In our approach,
the Knizknik-Zamolodchikov equations are the ground state
equations of the Hamiltonian (\ref{Sch})
and they provide a microscopic description
of the ``puncture" phase. Solving these equations
with a set of physical considerations
is tantamount to finding the conformal
blocks of a WZW theory.

For the spinning case, when $G = U(1)$,
one also has the same correspondence with
the $N=1$
super WZW theory up to a boundary condition for the fermionic components
of the superfields. Depending on the boundary condition,
one can have either the Neveu-Schwarz sector or the Ramond sector.
These possibilities follow from the fact that spinors can be double-valued
on the local coordinate patches of the 2-dimensional manifold.
It is known that
even at the quantum level,
super WZW theory is equivalent to the direct sum of a bosonic WZW theory
and a system of free Majorana fermions in the
adjoint representation of the gauge group \cite{Fuchs};
the spectrum of
supersymmetric WZW is just the bosonic WZW plus a number of
free fermions. Consequently, it is possible to interpret the solutions of
super Knizknik-Zamolodchikov equations as
the spinning non-abelian analogues of the Laughlin wavefunctions.
In particular, super $U(1)$ WZW with central charge $c = {3 \over 2}$
comprises of a compactifed boson and a free Majorana fermion, alias
Ising model at criticality. In this light,
the significance of the correlator of Ising model's energy operators
in FQHE \cite{MR}
becomes transparent. It ties up neatly with the spinning braid
group representation approach presented in section 4.2 where
the microscopic origin of the Moore-Read state was made manifest.

\newsubsection{FQHE is a manifestation of the ``puncture" phase}
\par
Our path integral representation
may leave an impression that
the many-body system in two dimensions is necessarily in the strongly
correlated phase.
The derivative $\dell{z} = \dell{z} - i A_z$, $A_z = 0$ is related to
$d_z \equiv \dell{z} - k \sum_j^N \frac{1}{z - w_j}$
by a {\em singular} gauge transformation:
\beq
A_{z} \longrightarrow A_z + \delldell{\varphi}{z} \, ,
\eeq
where
\beq
\varphi = - k \log \left[ (z - w_1) \cdots (z - w_N) \right] \, ,
\eeq
and $k \neq 0$. Except at a finite number of isolated points $w_j$,
the field strength is still zero (see (\ref{Gauss}));
$\mbox{\boldmath A} \equiv ( A_0, A_z, A_{\overline z} )$
is still a flat connection of a bundle over
$\mbox{\bf R}^2 - \{ w_1, \cdots, w_{N_h} \} $.
Mathematically, it seems that every free particle with a Hamiltonian
in the Schr\"odinger representation of the form
$ - \frac{2 \hbar^2}{m} \dell{z} \dell{\overline z}$ is gauge
equivalent to $ - \frac{\hbar^2}{m} \left(
d_z d_{\overline z} + d_{\overline z} d_z  \right)$.
If arbitrary singular gauge transformations are allowed,
the supposedly simply-connected configuration space becomes riddled with
punctures $w_j$ and thus {\em arbitrarily multiply-connected}.
Consequently, as section 2 shows, the wavefunction of the
free particle belongs to the representation space of the
braid group. In short, for arbitrary $k$,
every free particle or quasi-particle is always anyonic!

Certainly this is ostensible. It is {\em not} the picture we want to portray.
Under ordinary circumstances, the statistics of 2-dimensional
systems is still fermionic or bosonic.
As we have discussed in \cite{Lai-Ting}\cite{Ting-Lai},
the strongly correlated wavefunction is
a result of the configuration space being multiply
connected. Physically, this corresponds to the situation where
the system of particles is in a peculiar type of quantum phase wherein
each particle sees the others as punctures. Having understood
its origin, the next question is:
Why do the electrons see each other as punctures ?

In the experimentally verified
case of FQHE, plateaux develop only if the quality of the samples
is good, the temperature is at the vicinity of absolute zero, and the
magnetic field strength is strong.
Then, according to Laughlin's theory, the ground state of $N$ electrons
corresponding to a particular filling fraction is an incompressible
fluid. The quasi-excitation at a finite energy gap from the
ground state is characterized by fractional statistics.
When these conditions are not met, the collective
effect is absent and the statistics of the excitations
is just as usual;
the Hall conductance is not quantized with {\em fractional}
filling fraction\footnote{Of course, quantum Hall effect
with {\em integral} filling fractions can still occur.}. In other words,
one does not automatically have anyon (or braid group)
statistics for the excitations.

To understand how
the configuration space becomes multiply connected, we take the
illustrative analogy of Aharanov-B\"{o}hm effect. When
a 2-dimensional cross section is taken,
the infinitesimally thin solenoid appears as a puncture in the plane.
Hence,
our formulation can be applied to the Aharanov-B\"{o}hm effect as well;
(\ref{Lagrangian}) is the Lagrangian describing the phenomenon.
In the case of FQHE, we suspect that the high-frequency
cyclotron motion is the one that creates the puncture.
As is well known, each electron is in circular
motion in the presence of an uniform external magnetic field perpendicular
to the plane. When the field strength gets stronger,
the radius of the cirular motion
becomes smaller; the area enclosed by the circular motion vanishes when
the magnetic field strength is infinitely large. Other electrons cannot
stray into it anymore. Thus, the centre of the
circular motion becomes a puncture.
This is exactly the same as Aharanov-B\"{o}hm effect where
the interior of the solenoid is not accessible.
In some sense, the cyclotron motion with vanishing radius also chimes in
with the heuristic procedure of localizing or
attaching flux tubes onto the electrons. When the radius of the
cyclotron motion is not sufficiently small, the location of the
flux tube does not coincide with that of the electron. The
flux tube will sit on the electron only if the magnetic field is
strong enough to diminish the radius effectively to zero.
Therefore we see that the
{\em strong} external magnetic field is indispensable for
fractional quantum Hall effect with anyonic excitations
\footnote{The crucial role played by the background
magnetic field has also been discussed in \cite{GW}.
There, no-go theorems forbidding the existence of anyons with
any statistics on a torus is circumvented by the presence of
magnetic field.}.

On a more rigorous note,
the physical significance of the external magnetic field is reflected
in the mathematical requirement that any
{\em physical} wavefunction must be normalizable.
In this aspect,
the gauge potential of the magnetic field results in an exponential damping
term which renders the otherwise non-normalizable wavefunction of
an anyon normalizable.

Having assigned a bigger role for the external magnetic field in
FQHE, it is germane to speculate on the physical origin of the
threshold $B_0$ of incompressible excitation.
The experimental evidence of $B_0 > 0$ is built upon the
result of a systematic study of activation energies of the $p \over 3$
states on different specimens of comparable mobility,
with $p = 1, 2, 4, 5$ \cite{Stormer}. The data show that
the energy gaps vanish below 6 T.
We propose that the existence of $B_0$ may be understood in the
following manner.
Since the
formation of punctures is due to the {\em high}-frequency
cyclotron motion, there must be a minimum $\omega_0 = \frac{e B_0}{m^* c}$
below which the cyclotron motion fails to hem in and excise a small region
of the plane from being accessible to other electrons effectively.
In other words, below $\omega_0$, punctures are not formed and the
configuration space has trivial topology, and therefore no FQHE.
{}From this perspective, the integral quantum Hall effect is physically
distinct
from the FQHE. In the former case, each electron as a single particle
is very much indifferent to the existence of
its counterparts in the heterojunction. The FQHE differs fundamentally
in that it is the manifestation of strongly correlated
``puncture" phase.

\newsection{Summary}
\fpar
The braid group representation we have constructed is based on
the sum over the homotopically equivalent paths in the punctured
plane. The key element in our construction is to
employ the charged winding numbers to label the homotopy classes.
The homotopical constraint is then enforced through
the path-ordered Fourier integral. Naturally, information
about the non-simply connected configuration space is
translated into
the language of Hamiltonian associated with the path integral.
In this way, we explicitly show the link between non-abelian
anyon statistics and conformal field theory.
It is also of interest to point out the close relationship between
the braid group representation via path integal and
gauged non-linear Schr\"odinger equations \cite{referee}.

In this paper, we propose a quantization
procedure suitable for the construction of spinning
braid group representation. We find that super Knizhnik-Zamolodchikov
equations are the zero-energy equations of free spinning particles
when they see each other as punctures on the super plane. In other
words, if a system of spinning particles condenses in
the ``puncture" phase, the many-body ground state will be
characterized by the super Knizhnik-Zamolodchikov equations.
This is analogous to the spinless case which we addressed in previous
work \cite{Lai-Ting}. In a nutshell, everything boils down to
the quantum mechanical interpretation of (super) Kohno connection
as the topological constraint in terms of {\em charged} winding numbers.

We have applied the ``puncture" phase aspect of the representation theory
to the FQHE \cite{Ting-Lai}. Specifically,
spin-polarized wavefunctions constructed {\it a priori} by
Laughlin, and Moore and Read in the spinning case have been
shown to be the {\em exact} ground states of the respective
Hamiltonians (\ref{Ham}) and (\ref{spin Ham}). The repulsive
zero-range delta potentials in these Hamiltonians are consequent upon the
non-simply connected nature of the configuration space. This feature
agrees with the established views as reviewed by Laughlin and
Haldane in reference \cite{FQHE} where arguments involving numerical
studies and pseudopotential method are presented.
In addition, spin-singlet Halperin states describing unpolarized FQHE with
filling fractions $\frac{2}{2q + 1}$, $q = 2, 4, \cdots$
are also accountable in this framework. The common theme of all
these FQHE states is none other than the topology of the configuration
space.

The phenomenological implications of the braid group approach have
also been explored. We found that the energy gap of a FQHE state is
the zero-point energy of the quasi-excitation. Within the
interval of two Landau levels of the electrons,
sub-levels corresponding to the spectral signature of the
quasi-excitations exist (Figure 6). Indeed, the topological
excitations behave very much like electrons in the sense that
they also execute cyclotron motion under the influence of
the magnetic field. We have presented an experimentally testable
formula (\ref{ratio}) giving
the ratio of the excitation energy gaps of two Laughlin ground
states $|m_a \rangle, |m_b \rangle$.
Finally, the role of the magnetic field in the formation
of ``puncture" phase and hence the occurrence of threshold is
emphasized.

\vspace{1cm}
\noindent
We would like to thank Alwi, B. E. Baaquie, C. H. Oh and K. Singh for
discussions
on Witten's Chern-Simons theory \cite{Witten}\cite{CS-theory}\cite{GMM}.
C. T. would like to
thank DSO for indirect support in this work. He also wants to acknowledge
the hospitality of Prof Lim Hock and
the Laboratory for Image and Signal Processing at the National University
of Singapore. This work is dedicated to the memory of Heong-Moy Ting.

{\renewcommand{\Large}{\normalsize}
}

\begin{thebibliography}{99}
\bibitem{Tsui et al}
{D. Tsui, H. St\"{o}rmer and A. Gossard, Phys. Rev. Lett. {\bf 48}
(1982) 1559.}

\bibitem{Laughlin}
{ R. B. Laughlin, Phys. Rev. Lett. {\bf 50} (1983) 1395.}

\bibitem{Willett et al}
{R. L. Willett, J.P. Eisenstein, H. L. St\"{o}rmer, D. C. Tsui,
A. C. Gossard, and J. H. English, Phys. Rev. Lett. {\bf 59} (1987) 1776.}

\bibitem{Eisenstein et al}
{J.P. Eisenstein, R. L. Willett, H. L. St\"{o}rmer, D. C. Tsui,
A. C. Gossard, and J. H. English, Phys. Rev. Lett. {\bf 61} (1988) 997.}

\bibitem{HR}
{F. D. M. Haldane and E. H. Rezayi, Phys. Rev. Lett. {\bf 60} 956;
{\it ibid.} 1886. (1988)}

\bibitem{8/5}
{J.P. Eisenstein, H. L. St\"{o}rmer,
L. N. Pfeiffer and K. W. West,
Phys. Rev. Lett. {\bf 62} (1989) 1540.}

\bibitem{Lai-Ting}
{C. H. Lai and C. Ting, Phys. Lett. {\bf B 265} (1991) 341-346.}

\bibitem{Ting-Lai}
{C. Ting and C. H. Lai, Mod. Phys. Lett. {\bf B 5} (1991) 1293-1299.}

\bibitem{Wu}
{Y. S. Wu, Phys. Rev. Lett. {\bf 52} (1984) 2103-2106.}

\bibitem{spin}
{J. Fr\"ohlich and U. M. Studer, {\em U(1) $\times$ SU(2)-gauge invariance
of non-relativistic quantum mechanics, and generalized Hall effects},
ETH-TH/91-13; J. Fr\"ohlich, T. Kerler and P. A. Marchetti, {\em Non-abelian
bosonization in two-dimensional condensed matter physics},
DFPD 91/TH/16 (July 91).}

\bibitem{MR}
{G. Moore and N. Read, Nucl. Phys. {\bf B360}
(1991) 362-396.}

\bibitem{Witten}
{E. Witten, Comm. Math. Phys. {\bf 121} (1989) 351,
{\it ibid.} {\bf 137} (1991) 29.}

\bibitem{CS-theory}
{M. Bos and V. P. Nair,
Int. J. Mod. Phys. A{\bf 5} (1990) 959; H. Murayama, {\em Explicit
Quantization of the Chern-Simons Action}, University of Tokyo preprint,
UT-542, (Mar, 89); J. M. F. Labastida and A. V. Ramallo, Phys. Lett.
{\bf B227} (1989) 92; S. Elitzur, G. Moore, A. Schwimmer and N. Seiberg,
Nucl. Phys. {\bf B326} (1989) 108;
G. V. Dunne, R. Jackiw, and C. A. Trugenberger, Ann. Phys. {\bf 194} (1989)
197.}

\bibitem{GMM}
{E. Guadagnini, M. Martellini and M. Mintchev,
Nucl. Phys. {\bf B336} (1990) 581-609.}

\bibitem{KZ}
{V. Knizhnik and A. Zamolodchikov, Nucl. Phys. {\bf B247} (1984) 83.}

\bibitem{SWZW}
{P. di. Vecchia, V. Knizhnik, J. Petersen and P. Rossi, Nucl. Phys.
{\bf B253} (1985) 701.}

\bibitem{Fuchs}
{J. Fuchs, Nucl. Phys. {\bf B286} (1987) 455-484; Nucl. Phys. {\bf B318}
(1989) 631-654.}

\bibitem{Kohno}
{T. Kohno, Ann. Inst. Fourier, Grenoble, {\bf 37} (1987) 139;
Adv. Stud. Pure Math. {\bf 16} (1988) 255.}

\bibitem{FQHE}
{R. E. Prange and S. M. Girvin eds,
{\em The Quantum Hall Effect}, second edition, (Springer-Verlag, 1990).}

\bibitem{BEMS}
{A. P. Balachandran, E. Ercolessi, G. Morandi and A. M. Srivastava,
Int. J. Mod. Phys. {\bf B4} (1990) 2057-2196.}

\bibitem{Verlinde}
{E. Verlinde, {\em A Note on Braid statistics and the non-abelian
Aharonov-Bohm effect}, IASSNS-HEP-90/60 (Jan, 1991).}

\bibitem{Halperin}
{B. I. Halperin, Helv. Phys. Acta {\bf 56} (1983) 75;
Phys. Rev. Lett. {\bf 52} (1984) 1583.}

\bibitem{Stormer}
{For a review, see H. L. St\"ormer,
in {\em Frontiers in Physics, High Technology \& Mathematics},
edited by H. A. Cerderira and S. O. Lundqvist, World Scientific
(Singapore) (1990); G. S. Boebinger, A. M. Chang, H. L. St\"ormer
and D. C. Tsui, Phys. Rev. Lett. {\bf 55} (1985) 1606.}

\bibitem{Gepner-Witten}
{D. Gepner and E. Witten, Nucl. Phys. {\bf B278} (1986) 493-549.}

\bibitem{Tsuchiya-Kanie}
{A. Tsuchiya and Y. Kanie, Adv. Stud. Pure Math.
{\bf 16} (1988) 297;
Lett. Math. Phys. {\bf 13} (1987) 303.}

\bibitem{Fubini}
{S. Fubini, {\em Vertex operators and the quantum Hall effect},
CERN preprint TH-5763/90, (Nov, 90);
S. Fubini and C. A. L\"utken, Mod. Phys. Lett. {\bf A 6} (1991) 487-500.}

\bibitem{GW}
{M. Greiter and F. Wilczek, {\em Exact solutions and the adiabatic
heuristic for quantum Hall states}, IASSNS-HEP-91/45 (Jul, 91)}

\bibitem{referee}
{R. Jackiw and S.-Y. Pi,
Phys. Rev. Lett. {\bf 64} (1990) 2969;
S. M. Girvin, A. H. MacDonald, M. Fischer, S.-J. Rey and J. Sethna,
Phys. Rev. Lett. {\bf 65} (1990) 1671;
G. Dunne, R. Jackiw, S.-Y. Pi and C. Trugenberger,
Phys. Rev. {\bf D43} (1991) 1332.}

\end{thebibliography}
\end{document}